%% file: CWR-pilnu.tex
\RequirePackage{lineno}
\pdfoutput=1
\documentclass[preprint,prd,tightenlines, superscriptaddress]{revtex4}

\usepackage{graphicx} 
\usepackage{dcolumn}  
\usepackage{colordvi}
\usepackage{color}
\usepackage{epstopdf}
\usepackage{subfig}
\usepackage{caption}
\usepackage{amssymb}
\usepackage{amsmath}
\usepackage{url}
\graphicspath{{ps}}
\usepackage{hyperref}
\usepackage{tabularx}
\usepackage{comment}
\usepackage{booktabs}

\renewcommand{\arraystretch}{1.1}

\input belle2sym.tex

\def\Bsig        {\ensuremath{B_{\mathrm{sig}}}\xspace}

\def\X           {\ensuremath{X}\xspace}

\def \Y       {\ensuremath{\Upsilon(4S)}\xspace}

\begin{document}


\def\belletwo {{Belle II}\xspace}
\def\itbelletwo {{\it {Belle II}}\xspace}
\def\phaseiii {{Phase III}\xspace}
\def\itphaseiii {{\it {Phase III}}\xspace}

\newcommand\logten{\ensuremath{\log_{10}\;}}

\newcommand\bfrhoc{$\mathcal{B}$($B^0\to\rho^-\ell^+\nu_\ell$)}
\newcommand\bfrhoz{$\mathcal{B}$($B^+\to\rho^0\ell^+\nu_\ell$)}
\newcommand\bfpic{$\mathcal{B}$($B^0\to\pi^-\ell^+\nu_\ell$)}
\newcommand\bfpiz{$\mathcal{B}$($B^+\to\pi^0\ell^+\nu_\ell$)}


\def\lint {62.8 \invfb}
\def\procversion {{\it {proc11 + prompt}}\xspace}
\def\mcversion {{MC13a}\xspace}
\def\release {{\it {light-2002-ichep}}\xspace}
\def\feitraining {{\it {FEIv4\_2020\_MC13\_release\_04\_01\_01}}\xspace}
\def\sigprobcut {{$0.001$}\xspace}


\vspace*{-3\baselineskip}
\resizebox{!}{3cm}{\includegraphics{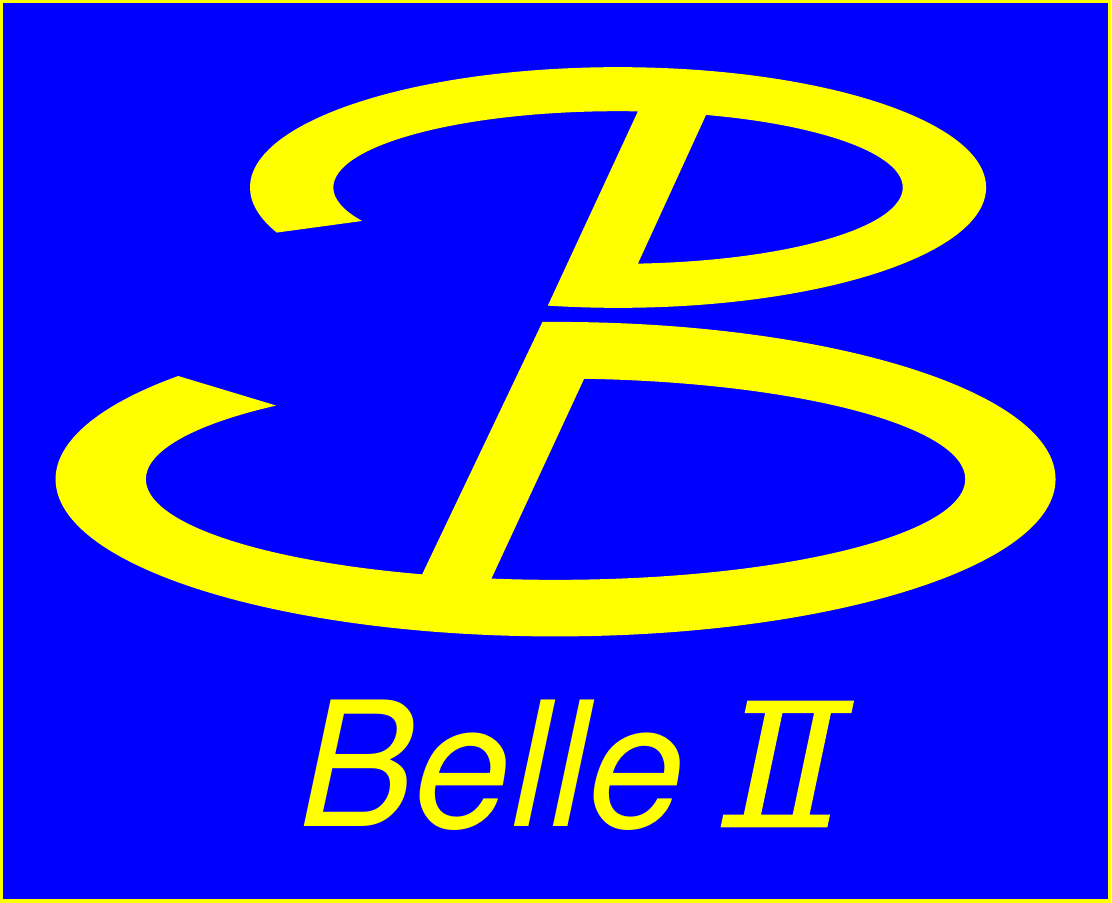}}

\vspace*{-5\baselineskip}
\begin{flushright}
\vspace{1em}
BELLE2-CONF-PH-2021-013\\
\today

\end{flushright}

\vspace{2em}
\title {\LARGE Exclusive $B \to X_u \ell \nu_\ell$ Decays with Hadronic Full-event-interpretation Tagging in 62.8\invfb of Belle II Data\\
\vspace{1em}
}

\input{authors-conf2021.tex}





\begin{abstract}

We present a reconstruction in early data of the semileptonic decay $B^+ \to \pi^0 \ell^+ \nu_\ell$, and first results of a reconstruction of the decays $B^+ \to \rho^0 \ell^+ \nu_\ell$ and $B^0 \to \rho^- \ell^+ \nu_\ell$ in a sample corresponding to 62.8\invfb of Belle II data using hadronic $B$-tagging via the full-event-interpretation algorithm. We determine the total branching fractions via fits to the distribution of the square of the missing mass, and find \bfpiz{} = (8.29 $\pm$ 1.99(stat) $\pm$ 0.46(syst)) $\times 10^{-5}$. We obtain $95\%$ CL upper limits on the branching fractions with \bfrhoc{} $ < 3.37 \times 10^{-4}$ and \bfrhoz{} $ < 19.7 \times 10^{-5}$. We also obtain an updated branching fraction for the $B^0 \to \pi^- \ell^+ \nu_\ell$ decay, \bfpic{} = (1.47 $\pm$ 0.29(stat) $\pm$ 0.05(syst)) $\times 10^{-4}$, based on the sum of the partial branching fractions in three bins of the squared momentum transfer to the leptonic system.


\end{abstract}


\maketitle

{\renewcommand{\thefootnote}{\fnsymbol{footnote}}}
\setcounter{footnote}{0}


\section{Introduction}

Since the start of its first physics operations in 2019, the Belle II detector has collected over 200\invfb of data from electron-positron collisions. These early data have been invaluable for investigating the performance of the detector and the analysis software.

In this paper, we present a reconstruction of the decays $B^0 \to \pi^- \ell^+ \nu_\ell$, $B^+ \to \pi^0 \ell^+ \nu_\ell$, $B^0 \to \rho^- \ell^+ \nu_\ell$ and $B^+ \to \rho^0 \ell^+ \nu_\ell$, where $\ell = e$, $\mu$, \footnote{Charge conjugate processes are implied for all quoted decays of $B$-mesons throughout this paper.} in a sample corresponding to 62.8\invfb of Belle II data via hadronic $B$-tagging provided by the full-event-interpretation (FEI) algorithm \cite{Keck:2018lcd}. These decays are considered golden modes for precise determinations of the magnitude of the Cabibbo-Kobayashi-Maskawa (CKM) matrix element $|V_{\mathrm{ub}}|$. Whilst the integrated luminosity collected at present is too small to provide a competitive measurement, we demonstrate the first steps towards extracting a measurement of $|V_{\mathrm{ub}}|$ using $B^0 \to \pi^- \ell^+ \nu_\ell$ decays.

\section{The Belle II Detector}
The Belle II detector is described in detail in Ref. \cite{Abe:2010sj}. The innermost layers are known collectively as the vertex detector, or VXD, and are dedicated to the tracking of charged particles and the precise determination of particle decay vertices. The VXD is composed of two layers of silicon pixel sensors surrounded by four layers of silicon strip detectors. The central drift chamber (CDC) surrounds the VXD, encompassing the barrel region of the detector, and is primarily responsible for the reconstruction of charged particles and the determination of their momenta and electric charge.

Particle identification is provided by two independent Cherenkov-imaging instruments, the time-of-propagation
counter and the aerogel ring-imaging Cherenkov detector, located in the barrel and forward endcap regions of the detector, respectively. The electromagnetic calorimeter (ECL) encases all of the previous layers and is used primarily for the determination of the energies of charged and neutral particles. A superconducting solenoid surrounds the inner components and provides the 1.5 T magnetic field required by the various sub-detectors. Finally, the $K^0_L$- and muon detector forms the outermost detector layer aimed at the detection of $K^0_L$ mesons and muons.

\section{Data sets}
\label{sec:data sets}

The amount of data studied for this analysis corresponds to an integrated luminosity of 62.8\invfb. To simulate signal and background, fully simulated Monte Carlo (MC) samples of decays of pairs of charged or neutral $B$ mesons, as well as 
continuum \epem \to \qqbar ($q = u,\,d,\,s,\,c$) processes are used, generated alongside beam background effects including beam scattering and radiative processes. Table \ref{table:MC13} lists the number of events used for each of the MC components. For the analysis of $B\to\pi\ell\nu_\ell$ decays, MC samples corresponding to a total integrated luminosity of 200\invfb are used, with 400\invfb equivalent samples used for $B\to\rho\ell\nu_\ell$ studies. 

\begin{table} [h!]
\caption{Simulated event yields used for the analyses, equivalent to 200(400)\invfb for $B\to\pi\ell\nu_\ell$ and $B\to\rho\ell\nu_\ell$, respectively.}\label{table:MC13}
\begin{tabular}{ccc}
\hline\hline
&   \multicolumn{2}{c}{$N_{\mathrm{events}}$ ($\times 10^6$)} \\
& $B\to\pi\ell\nu_\ell$ analysis & $B\to\rho\ell\nu_\ell$ analysis\\ 
\hline
 $B^+B^-$ &  108.0 & 216.0\\ 
 $B^0\bar{B}^0$ &  102.0 & 204.0\\ 
 $c\bar{c}$ &  265.8 & 531.6\\ 
 $u\bar{u}$ &  321.0 & 642.0\\ 
 $s\bar{s}$ &  76.6 & 153.2\\ 
 $d\bar{d}$ &  80.2 & 160.4\\ 
 \hline\hline
\end{tabular}
\end{table}

In addition to the generic MC samples, dedicated samples of $B \to \X_u \ell \nu_\ell$ decays, where $X_u$ is a hadronic system resulting from the quark flavor transition $b \to u$, are used to model signal decays and related backgrounds. The $X_u$ system includes both resonant and non-resonant contributions using the hybrid modelling technique of Ref. \cite{Ramirez:1990db}, which is briefly described here.

Each $B^+ \to \X_u \ell \nu_\ell$ and $B^0 \to \X_u \ell \nu_\ell$ sample consists of a total of 50 million resonant (R) events containing the relevant exclusive decays as well as 50 million non-resonant (I) events corresponding to the inclusive component, simulated using the BLNP heavy-quark-effective-theory-based model \cite{Lange:2005ll}. These samples are then combined together and the eFFORT tool \cite{markus_prim_2020_3965699} is used to calculate an event-wise weight $w_i$ in three-dimensional bins of the generated lepton energy in the $B$-frame, $E^B_\ell$, the squared four-momentum transfer to the leptonic system, $q^2$, and the mass of the hadronic system containing an up-quark, $M_X$, such that $H_i = R_i + w_i I_i.$ The number of total hybrid events per bin, $H_i$, is the sum of the number of resonant events $R_i$ and the number of inclusive events $I_i$ scaled down by the appropriate weight $w_i$. 

The $B \to \X_u \ell \nu_\ell$ events from the generic 200(400)\invfb MC samples are replaced with the equivalent amount of this hybrid re-weighted MC for the $B\to\pi\ell\nu_\ell$ and $B\to\rho\ell\nu_\ell$ analyses, respectively.

\section{full event interpretation}
\label{sec:FEISkim}
\begin{figure}[h!]
\begin{center}
\includegraphics[scale=0.8]{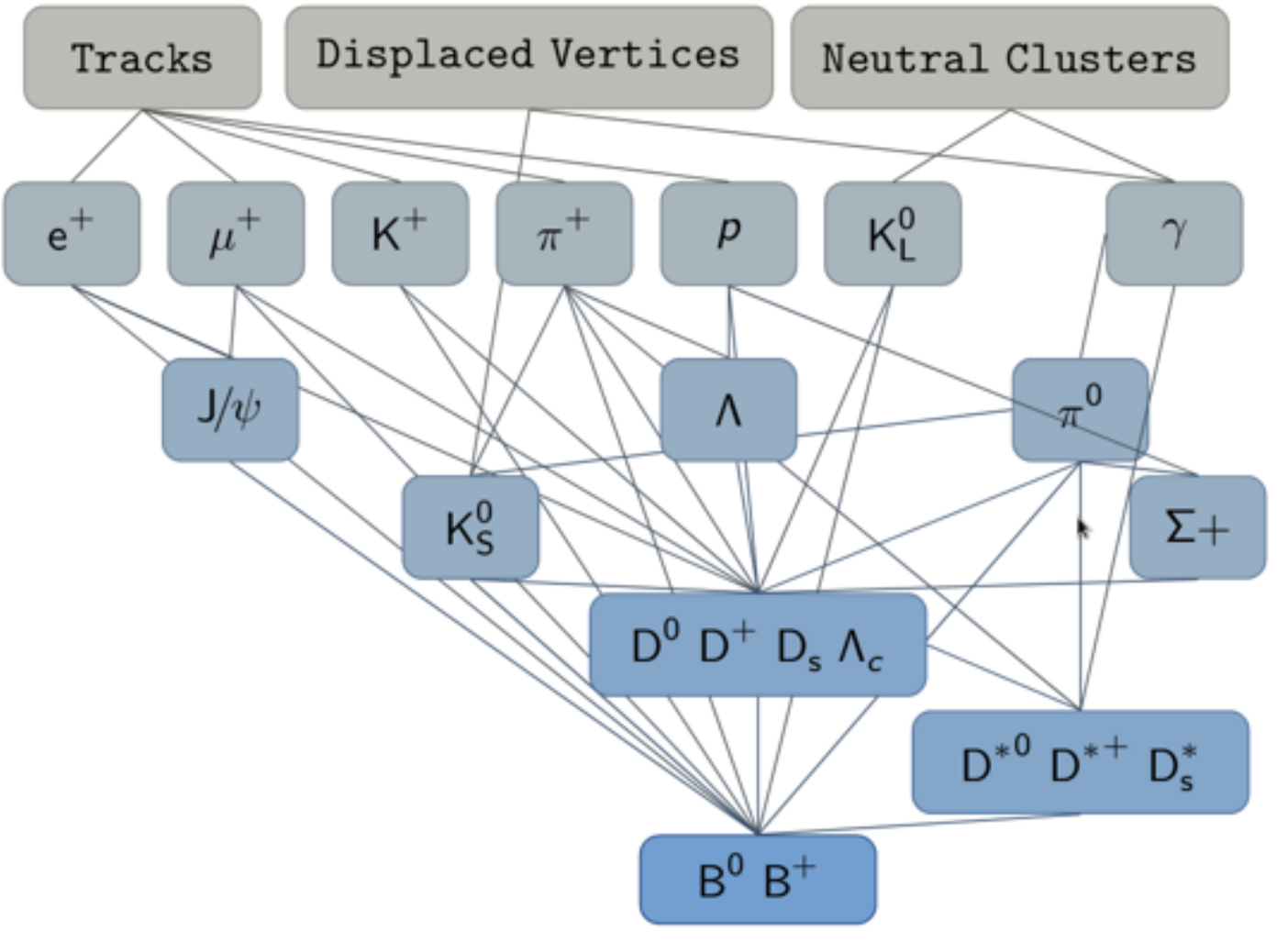}
  \caption{Hierarchical structure of the full-event-interpretation tagging algorithm. 
    }
  \label{fig:fei}
\end{center}
\end{figure}

The second $B$-meson in the $B\overline{B}$ pair is reconstructed using the Full Event Interpretation (FEI) algorithm \cite{Keck:2018lcd} to tag the event.
The FEI is a machine learning algorithm developed for $B$-tagged analysis at Belle II. It supports both hadronic and semileptonic tagging, reconstructing $B$ mesons across more than 4000 individual decay chains. The algorithm utilises a FastBDT software package that trains a series of multi-variate classifiers for each tagging channel via a number of stochastic gradient-boosted decision trees \cite{Keck:2016tk}. The training is performed in a hierarchical manner with final-state particles being reconstructed first from detector information. The decay channels are then built up from these particles as illustrated in Figure \ref{fig:fei}, with the reconstruction of the $B$-mesons performed last. For each $B$-meson tag candidate reconstructed by the FEI, a value of the final multi-variate classifier output, the \texttt{SignalProbability}, is assigned. The \texttt{SignalProbability} is distributed between zero and one, representing candidates identified as being background-like and signal-like, respectively.

FEI skims of both data and MC are produced centrally by the Belle II collaboration, and are available for use in analyses. These include both hadronic and semileptonic skims, and involve the application of the FEI together with a number of loose selections that aim to reduce the sample sizes with little to no loss of signal events.

For the hadronic FEI, the minimum number of tracks per event satisfying certain quality criteria is set to three. The vast majority of $B$-meson decay chains corresponding to the hadronic FEI channels include at least three charged particles, and such a criterion is useful at suppressing background from non-$B\bar{B}$ events. Requirements are placed on the track parameters as defined in \cite{belle2tracking:2021} to ensure close proximity to the interaction point (IP), with the distance from the center of the detector along the $z$-axis (corresponding to the solenoidal-field axis) and in the transverse plane satisfying $|z_0| < 2.0\,\rm cm$ and $|d_0| < 0.5\,\rm cm$, respectively. A minimum threshold $p_t >$ 0.1 GeV/$c$ is placed on the particle transverse momentum. Similar restrictions are applied to the ECL clusters in the event, with at least three clusters required within the polar angle acceptance of the CDC, 0.297 $< \theta <$ 2.618, that satisfy a minimum energy threshold $E > 0.1$ GeV. The total detected energy per event is required to be at least 4 GeV. The total energy deposited in the ECL is restricted to $2\,{\rm GeV} < E_{\rm ECL} < 7\,{\rm GeV}$, however, to suppress events with an excess of energy deposits due to beam background.

The FEI typically results in many $B_{\mathrm{tag}}$ candidates per event. The number of these candidates is reduced with selections on the beam-constrained mass, $M_{\mathrm{bc}}$, and energy difference, $|\Delta E|$,\\

$$M_{\mathrm{bc}} = \sqrt{\frac{E_{\mathrm{beam}}^{2}}{4}-\vec{p}^{\,2}_{B_{\mathrm{tag}}}}\hspace{0.5em},  \hspace{3em} \Delta E = E_{B_{\mathrm{tag}}} - \frac{E_{\mathrm{beam}}}{2},$$
where $E_{\mathrm{beam}}$ is the centre-of-mass (CMS) energy of the $e^+e^-$ system, 10.58 GeV, and $\vec{p}_{B_{\mathrm{tag}}}$ and $E_{B_{\mathrm{tag}}}$ are the  $B_{\mathrm{tag}}$ momentum and energy in the CMS frame, respectively. The criteria applied during the hadronic FEI skim are $M_{\mathrm{bc}} > 5.24$ GeV/$c^2$ and $|\Delta E| < 0.2$ GeV.

Finally, a loose requirement on the $B_{\mathrm{tag}}$ classifier output,  \texttt{SignalProbability} $> 0.001$, provides further background rejection with little signal loss. 

\section{Event Selection and Analysis Strategy}
\label{sec:selections}

For this analysis, the distribution of the square of the missing mass, $M_{\mathrm{miss}}^2$, is the variable chosen for the determination of the data yields. We define the four-momentum of the signal $B$-meson \Bsig{} in the CMS frame as follows,

\begin{equation*}
    p_{B_{\mathrm{sig}}} \equiv  (E_{B_{\mathrm{sig}}}, \vec{p}_{B_{\mathrm{sig}}}) = \left(\frac{m_{\Upsilon(4S)}}{2}, -\vec{p}_{B_{\mathrm{tag}}}\right),
\end{equation*}
where $m_{\Upsilon(4S)}$ is the known $\Upsilon$(4S) mass \cite{Zyla:2020zbs}. We set the energy of $B_{\mathrm{sig}}$ to be half of the $\Upsilon$(4S) rest mass, and take the $B_{\mathrm{sig}}$ momentum to be the negative $B_{\mathrm{tag}}$ momentum. We then define the missing four-momentum as

\begin{equation*}
    p_{\mathrm{miss}} \equiv (E_{\mathrm{miss}}, \vec{p}_{\mathrm{miss}}) =  p_{B_{\mathrm{sig}}} - p_Y,
\end{equation*}
where $Y$ represents the combined lepton-hadron (pion or $\rho$-meson) system. The square of the missing momentum can then simply be defined as $M_{\mathrm{miss}}^2 \equiv p_{\mathrm{miss}}^2$.

The event selections applied follow closely those from a 2013 study \cite{Sibidanov:2013sb} of exclusive, hadronically-tagged $B \to \X_u \ell \nu_\ell$ decays reconstructed in the full 711\invfb Belle data set. All selections are applied in addition to the hadronic FEI skim criteria detailed in the previous section.

At the event-level, a loose selection on the second normalised Fox-Wolfram moment \cite{Wolfram:1978fw} is applied for the $B\to\pi\ell\nu_\ell$ analysis, R2 $<$ 0.4, in order to suppress events from the continuum background. An alternative approach is taken for reconstructed $B\to\rho\ell\nu_\ell$ candidates due to the larger continuum background contribution in this channel. Here several variables including the second normalised Fox-Wolfram moment, $B$-meson thrust angles and magnitudes, CleoCones \cite{Cleo:1996} and modified Fox-Wolfram moments \cite{Bfactories:2014} are combined into a boosted decision tree classifier, again from the FastBDT software package. Out of the 64 available variables, the 44 with the highest classification power are selected. After applying the multivariate method to all $\Upsilon$(4S) candidates, a classifier requirement is chosen such that 94.1\% of continuum events are rejected and 83.5\% of signal events are retained in $B^+ \to \rho^0 \ell \nu_\ell$ candidates, and  96.1\% of continuum events are rejected and 84.8\% of signal events are retained in $B^0 \to \rho^+ \ell \nu_\ell$ candidates.

To reject incorrectly reconstructed $B_{\mathrm{tag}}$ candidates, the tag-side beam-constrained mass criterion is tightened to $M_{\mathrm{bc}} > 5.27 \text{GeV}/c^2$. The $B_{\mathrm{tag}}$ candidate having the highest value of the \texttt{SignalProbability} classifier output is retained in each event.

For the reconstructed electrons and muons, track impact parameters are used to select tracks originating close to the interaction point, thereby suppressing background events from beam scattering and radiative effects. Tracks are required to have $z$-axis and transverse-plane distances from the IP of $|dz| < 5\,{\rm cm}$ and $dr < 2$ cm, respectively. 
Only those leptons within the acceptance of the CDC are selected. Electrons and muons are identified through selection criteria on the particle identification variables provided by the Belle II analysis software framework \cite{Kuhr:B2}. These variables describe the probability that each species of charged particle generates the particle-identification signal observed, and are built from a combination of the information returned from all of the individual sub-detectors except the SVD. Electron and muon candidates are each required to have an identification probability above $0.9$ as assigned by the appropriate reconstruction algorithm.
For $B\to\pi\ell\nu_\ell$ decays, a minimum threshold on the lab-frame momentum is placed on the reconstructed leptons, with $p_{\mathrm{lab}} > 0.3$ GeV/$c$ for electrons and $p_{\mathrm{lab}} > 0.6$ GeV/$c$ for muons. A single threshold of $p_{\mathrm{lab}} > 0.4$ GeV/$c$ is used for both electrons and muons to reconstruct $B\to\rho\ell\nu_\ell$ decays. 

To suppress leptons originating from $\gamma$ conversions in the detector as well as $J/\psi$ and $\psi '$ decays, the invariant mass of all oppositely charged lepton pairs is required to exceed 0.1 GeV/$c^2$ and be outside the 3.00 -- 3.12 GeV/$c^2$ and 3.6 -- 3.75 GeV/$c^2$ mass ranges in the $B\to\rho\ell\nu_\ell$ part of the analysis.

The four-momenta of the reconstructed electrons are also corrected in order to account for bremsstrahlung radiation. The correction methods employed differ slightly between the $B\to\pi\ell\nu_\ell$ and $B\to\rho\ell\nu_\ell$ analyses. For reconstructed $B\to\pi\ell\nu_\ell$ decays, any energy deposit in the ECL not associated with a track is considered a bremsstrahlung photon if it is detected within an angle of 3$^{\circ}$ from a reconstructed electron candidate. In these cases, the four-momentum of the photon is added to the electron, and the photon is excluded from the rest of the event. If multiple photons meet this criteria, only the photon nearest the electron candidate is considered.

For the $B\to\rho\ell\nu_\ell$ case the same method is used but the electron candidates are separated into three momentum regions, from 0.4 to 0.6 GeV/$c$, from 0.6 to 1.0 GeV/$c$ and above 1.0 GeV/$c$. For the first region, no corrections are applied. For the second region, the energy of all photons with an energy below 600 MeV and within 4.8$^{\circ}$ of the electron candidate is added to the candidate. In the third region, the angular threshold is tightened to 3.4$^{\circ}$ while the photon energy selection is relaxed to 1.0 GeV. These values are determined by minimizing the root mean square of the difference between generated and reconstructed electron momenta.

Finally, a single lepton is kept per event with the highest value of the lepton identification probability as described above.

For the reconstructed charged pions, similar impact parameter criteria are applied as those for the leptons, with $dr < 2$ cm and $|dz| < 4$ cm. Similarly, the charged pion tracks are only selected within the CDC acceptance. A selection on the relevant particle identification variable is also applied, with a particle identification probability above $0.6$. The sign of the charge of the reconstructed pion is explicitly required to be opposite that of the lepton for the $B^0 \to \pi^- \ell^+ \nu_\ell$ case.

In reconstructing neutral pions, different thresholds on the photon-daughter energies are required, depending on the polar direction of the candidate photon. These requirements are $E > 0.080$ GeV for the forward end-cap, $E > 0.030$ GeV for the barrel region and $E > 0.060$ GeV for the backward end-cap. A selection on the diphoton mass is also implemented, with 0.120   $\text{GeV}/c^2  < M_{\gamma\gamma} < 0.145$ $\text{GeV}/c^2$. For $B^+ \to \pi^0 \ell^+ \nu_\ell$ candidates, a selection on the cosine of the lab-frame opening angle of the $\pi^0$ photon daughters is also applied in order to reject backgrounds from photon pairs that do not originate from $\pi^0$ decays, cos$\psi_{\gamma\gamma} > 0.25$. For the $\pi^0$ candidates used to reconstruct charged $\rho$-mesons, a tighter selection is applied, with  cos$\psi_{\gamma\gamma} > 0.4$.

To reconstruct $\rho$-mesons, either two charged pions or a charged pion and a neutral pion are combined. The invariant mass of this pion pair is required to be between 0.333 GeV/$c^2$ and 1.217 GeV/$c^2$ to reject contributions from non-resonant $B \to \pi \pi \ell \nu_\ell$ decays. If multiple charged $\rho$-meson candidates are reconstructed, the candidate with the highest energy in the centre-of-mass frame is chosen.

The centre-of-mass frame four-momenta of the reconstructed meson and lepton are combined into the system $Y$. The angle between the flight directions of the signal $B$-meson as inferred from initial beam conditions and the $Y$ is then used to select events more likely to originate from the decay of interest. The cosine of this angle, $\mathrm{cos}(\theta_{BY})$, is defined as
$$
\mathrm{cos}(\theta_{BY}) = \frac{2E_{\mathrm{beam}}E_Y - m_{B_{\mathrm{sig}}}^2 - m_Y^2}{2|\vec{p}_{B_{\mathrm{sig}}}||\vec{p}_Y|},
$$
where $m_{B_{\mathrm{sig}}}$ is the invariant mass of the signal $B$-meson and $E_Y$, $m_Y$ and $\vec{p}_Y$ are the energy, invariant mass and momentum of the $Y$ system, respectively. A value of $|\mathrm{cos}(\theta_{BY})| < 1$ is expected if only a neutrino is missing in the reconstruction. However, to account for resolution effects and to avoid introducing potential bias in the background $M_{\mathrm{miss}}^2$ distributions, this requirement is loosened to $|\mathrm{cos}(\theta_{BY})| < 3$.

To ensure that the reconstructed lepton and pion tracks originate from the same vertex in $B^0 \to \pi^- \ell^+ \nu_\ell$ decays, the difference between the $z$-coordinates of both tracks is required to be $|z_\ell - z_\pi| < 1$ mm.

For this analysis, a single $\Upsilon$(4S) candidate with the lowest value of $M_{\mathrm{miss}}^2$ is retained per event. A minimum threshold on the missing energy, $E_{\mathrm{miss}}$, is placed to account for the neutrino, with $E_{\mathrm{miss}} > 0.3$ GeV. All remaining tracks and clusters in the ECL after the reconstruction of the $\Upsilon$(4S) are combined into a single system known as the rest-of-event (ROE). Events in which additional tracks satisfying the conditions $dr < 2$ cm, $|dz| < 5$ cm and $p_t > 0.2$ GeV/$c$ remain after the reconstruction of the $\Upsilon$(4S) are excluded. The CMS energies in the ROE corresponding to deposits of neutral particles in the ECL are summed for those clusters satisfying energy selection criteria of $E > 0.1$ GeV, $E > 0.09$ GeV and $E > 0.16$ GeV for the forward end-cap, barrel and backward end-cap regions, respectively. This extra energy is required to be below a maximum value of 
$E_{\mathrm{residual}} < 1.0$ GeV for $B^0 \to \pi^- \ell^+ \nu_\ell$, $E_{\mathrm{residual}} < 0.6$ GeV for $B^+ \to \pi^0 \ell^+ \nu_\ell$, and $E_{\mathrm{residual}} < 0.7$ GeV for $B\to\rho\ell\nu_\ell$ candidates.




\section{Results}
\label{sec:results}

\begin{figure}[h!]
\begin{center}
\includegraphics[scale=0.38]{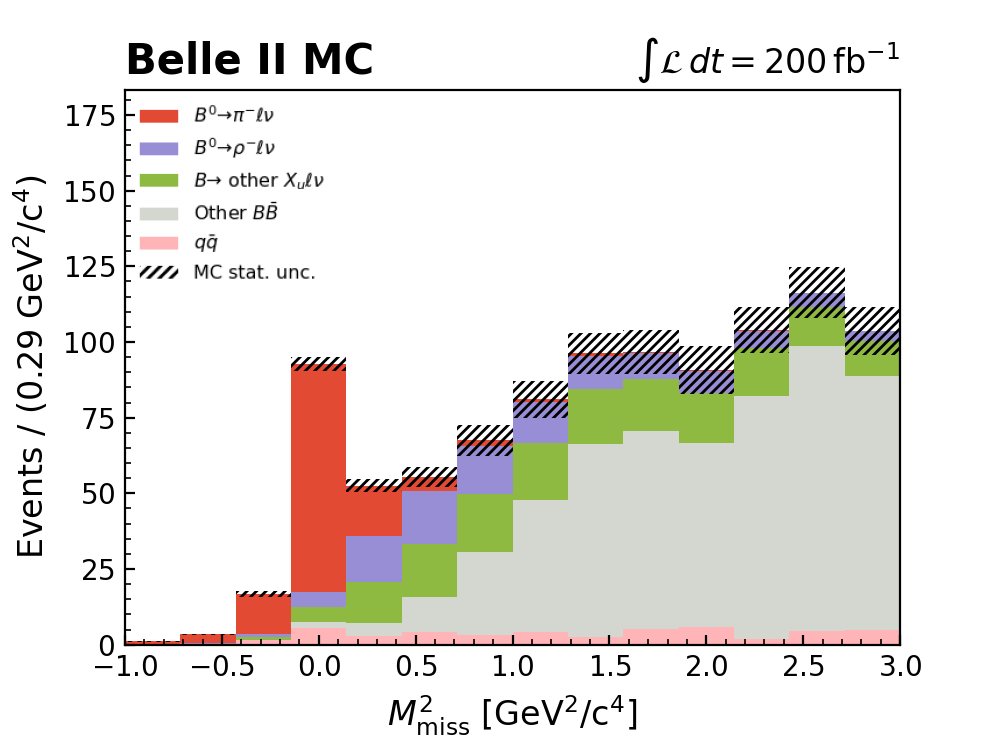}
\includegraphics[scale=0.38]{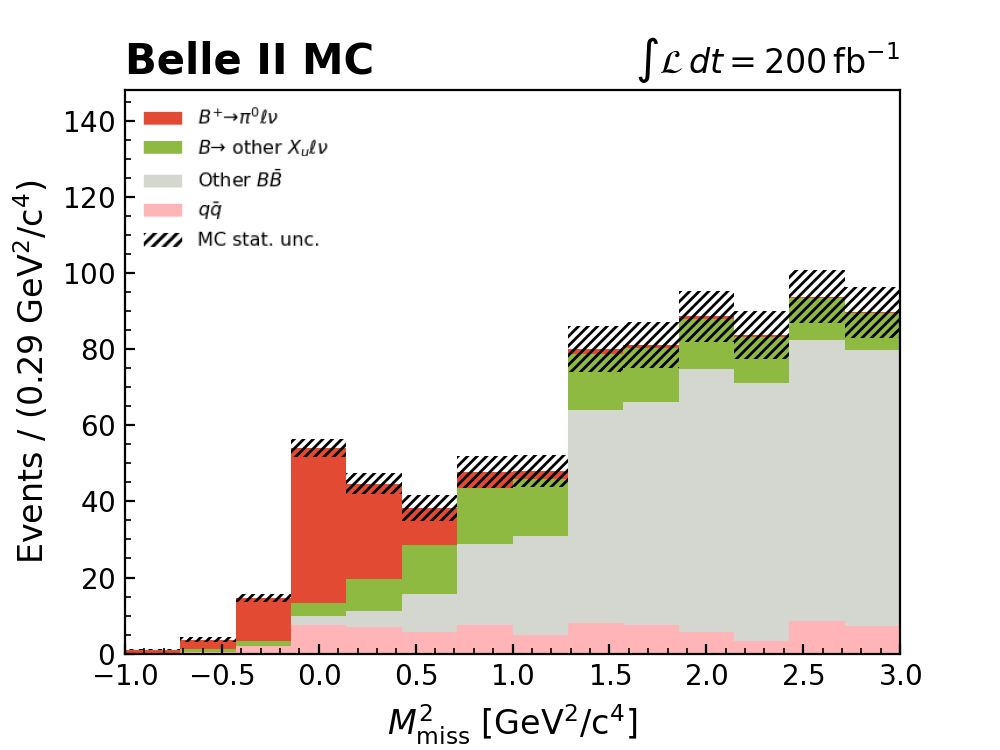}
  \caption{$M_{\mathrm{miss}}^2$ distributions for (left) $B^0 \to \pi^-\ell^+\nu_\ell$ and (right) $B^+ \to \pi^0\ell^+\nu_\ell$ candidates reconstructed from a sample corresponding to 200\invfb of simulated data. The branching fractions of the signal components are normalised to be consistent with the world averages from Ref. \cite{Zyla:2020zbs}.}
  \label{fig:prefitpilnu}
\end{center}
\end{figure}

\begin{figure}[h!]
\begin{center}
\includegraphics[scale=0.6]{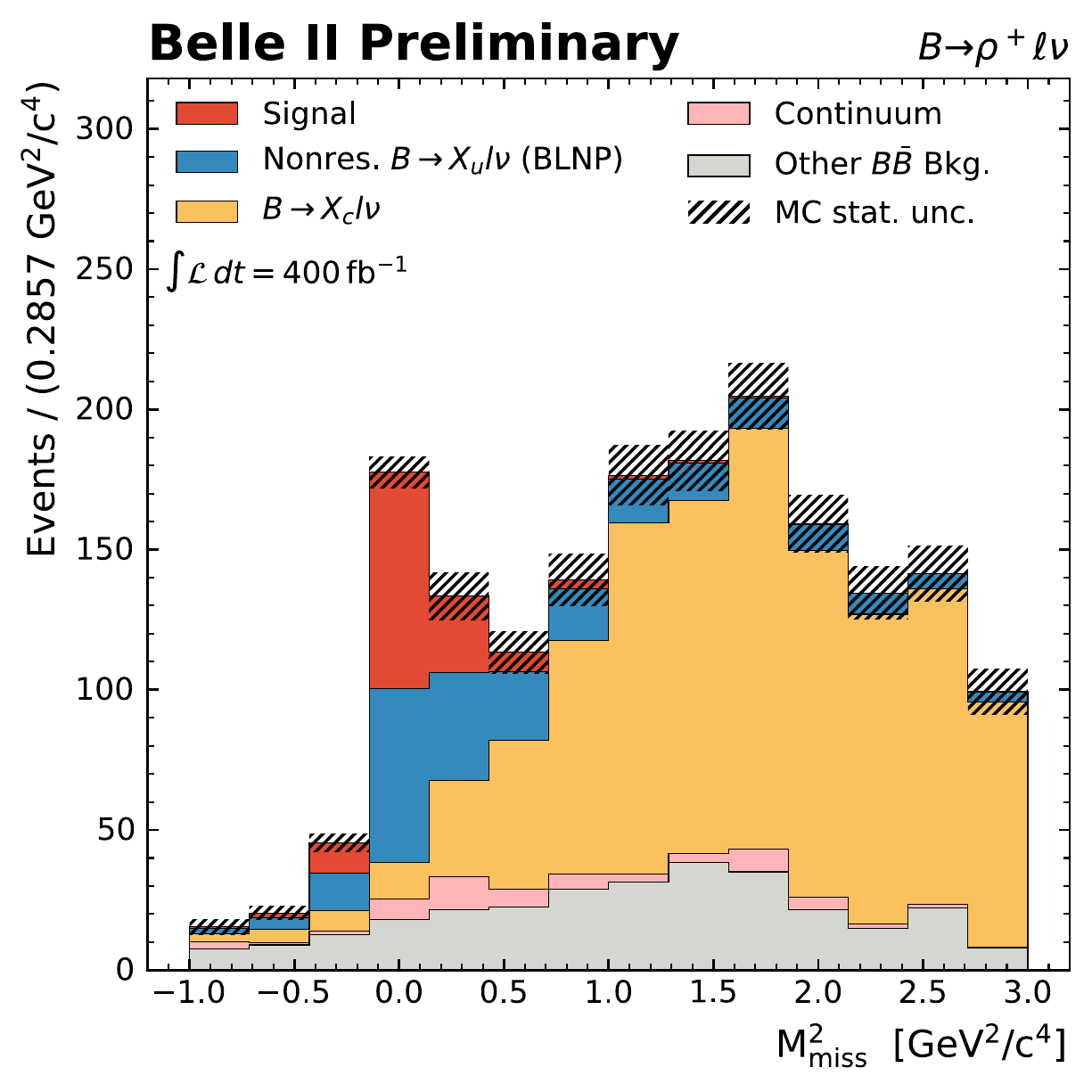}
\includegraphics[scale=0.6]{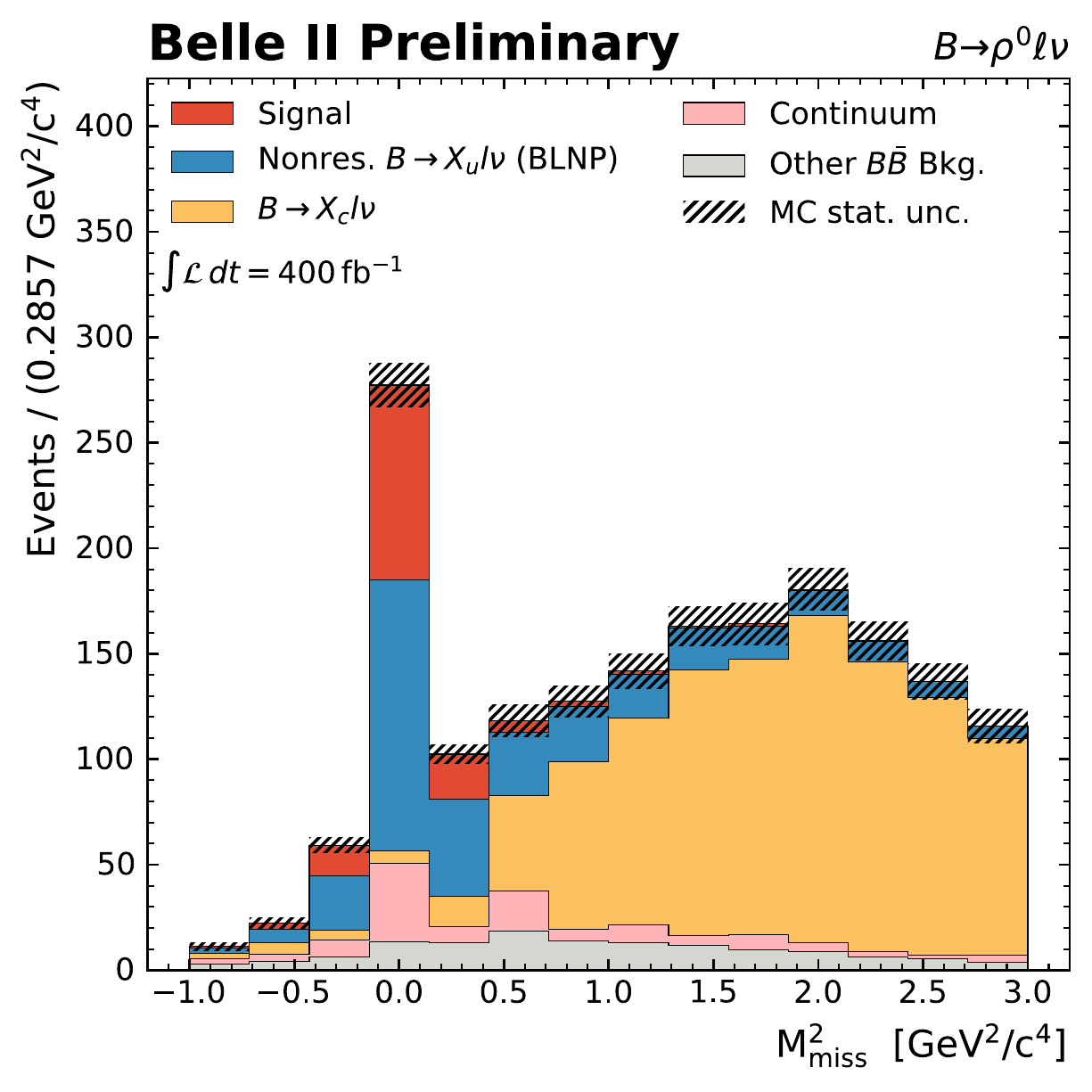}
  \caption{$M_{\mathrm{miss}}^2$ distributions for (left) $B^0 \to \rho^-\ell^+\nu_\ell$ and (right) $B^+ \to \rho^0\ell^+\nu_\ell$ candidates reconstructed from a sample corresponding to 400\invfb of simulated data. The branching fractions of the signal components are normalised to the world averages from Ref. \cite{Zyla:2020zbs}.}
  \label{fig:prefitrholnu}
\end{center}
\end{figure}

The resultant $M_{\mathrm{miss}}^2$ distributions in MC after all analysis selections are displayed in Figures \ref{fig:prefitpilnu} and \ref{fig:prefitrholnu} for $B\to\pi\ell\nu_\ell$ and $B\to\rho\ell\nu_\ell$ decays, respectively. In these distributions, the MC has been separated into distinct components to illustrate the relative contributions of various background processes. For the $B\to\pi\ell\nu_\ell$ analysis, these include the cross-feeds from $B^0 \to \rho^- \ell^+ \nu_\ell$ and other $B \to X_u \ell \nu$ decays, as well as candidates reconstructed from other generic $B\bar{B}$ and continuum events. 
For the $B\to\rho\ell\nu_\ell$ analysis, the background distributions shown in Figure \ref{fig:prefitrholnu} include the cross-feeds from non-resonant $B^0 \to \pi \pi \ell^+ \nu_\ell$ decays, a large contribution from $B \to X_c \ell \nu_\ell$ decays as well as candidates reconstructed from other generic $B\bar{B}$ and continuum events. 

A number of corrections and scaling factors are applied to the simulated data used for Figures \ref{fig:prefitpilnu} and \ref{fig:prefitrholnu}. For all reconstructed modes, the total number of MC events is scaled down by a hadronic FEI calibration factor of $0.79 \pm 0.02$ for events reconstructed using neutral $B$-meson tags, and $0.63 \pm 0.02$ for those reconstructed via charged $B$-meson tags. These factors are applied in order to account for the differences in the tag-side reconstruction efficiency of the FEI between data and MC. An independent study is performed in order to evaluate these factors through fitting the lepton momentum spectrum in $B\to X \ell\nu_\ell$ decays and taking the ratio of signal events in data and MC. For decays involving neutral $\pi^0$ mesons, namely $B^+ \to \pi^0\ell^+\nu_\ell$ and $B^0 \to \rho^-\ell^+\nu_\ell$, an additional scaling factor SF$_{\pi^0} = 0.945 \pm 0.041$ is also applied to the total MC component to correct for differences in the $\pi^0$ reconstruction efficiency between MC and data. This factor is also determined via an independent study of $\eta \to 3\pi^0$ decays, in which the ratio of signal events in data and MC is determined through fitting the invariant mass of the reconstructed $\eta$ meson.

Furthermore, each MC component is weighted by a set of corrections to account for the differences in the lepton identification efficiencies and the pion and kaon fake-rates between MC and data. These corrections are obtained in an independent study \cite{LeptonID:2318} and are evaluated per event based on the magnitude of the lab-frame momentum $p$ and polar angle $\theta$ of the reconstructed lepton tracks. For the $B^0 \to \pi^-\ell^+\nu_\ell$ and $B\to \rho\ell\nu_\ell$ decays, a similar set of MC corrections are applied for the charged pion identification efficiencies and the fake-rates due to charged kaons.

The event selection criteria are applied to data along with some data-specific correction factors. Charged particle momenta are multiplied by a factor 1.00056 to correct for momentum-scale differences between data and MC. For reconstructed $B^+ \to \pi^0\ell^+\nu_\ell$ decays in data, an additional correction is applied to scale the energies of the photons assigned to the signal $\pi^0$ candidates to account for known photon energy biases.

\begin{figure}[h!]
\begin{center}
\includegraphics[scale=0.38]{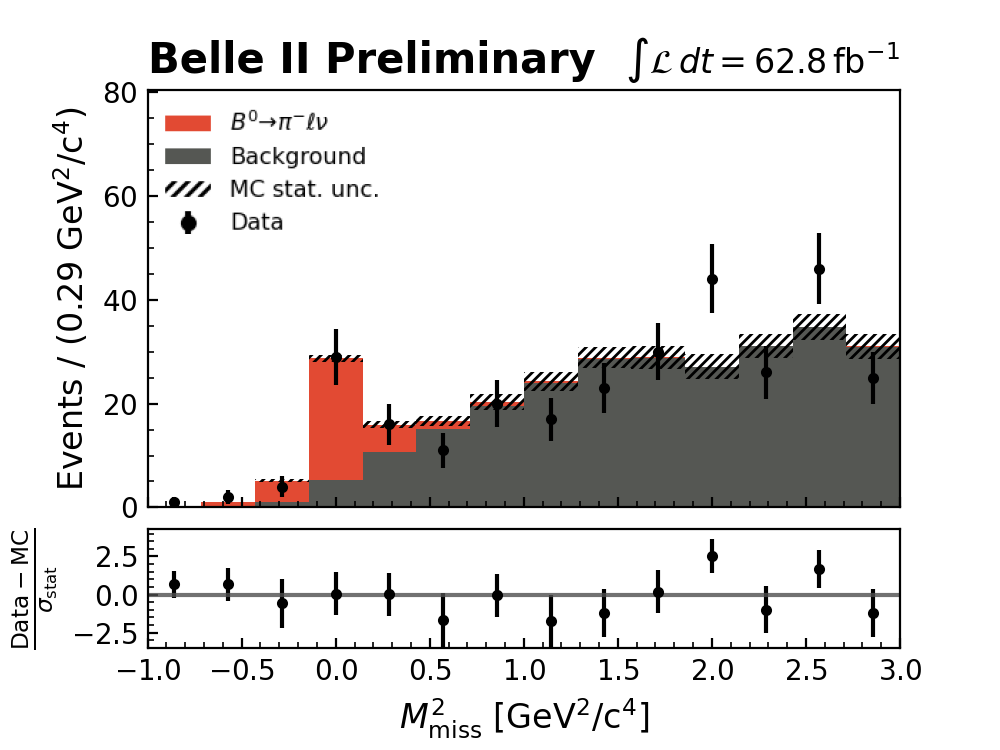}
\includegraphics[scale=0.38]{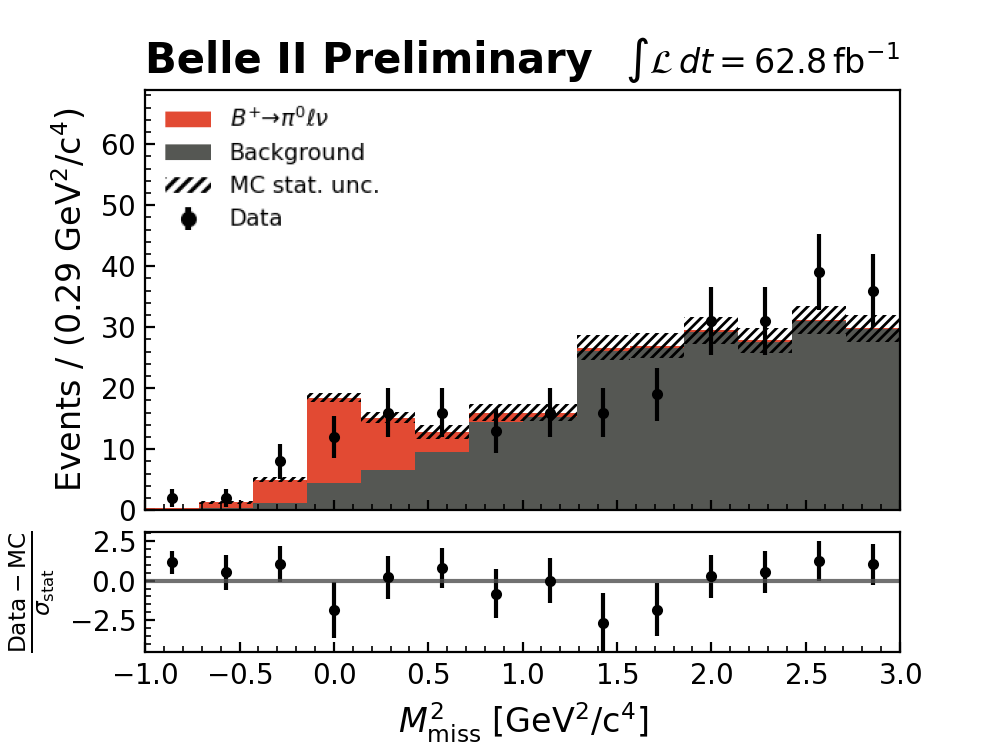}
  \caption{$M_{\mathrm{miss}}^2$ distributions for (left) $B^0 \to \pi^-\ell^+\nu_\ell$ and (right) $B^+ \to \pi^0\ell^+\nu_\ell$ candidates reconstructed from a sample corresponding to 62.8 \invfb of experimental data with fit projections overlaid.}
  \label{fig:postfitpilnu}
\end{center}
\end{figure}

\begin{figure}[h!]
\begin{center}
\includegraphics[scale=0.6]{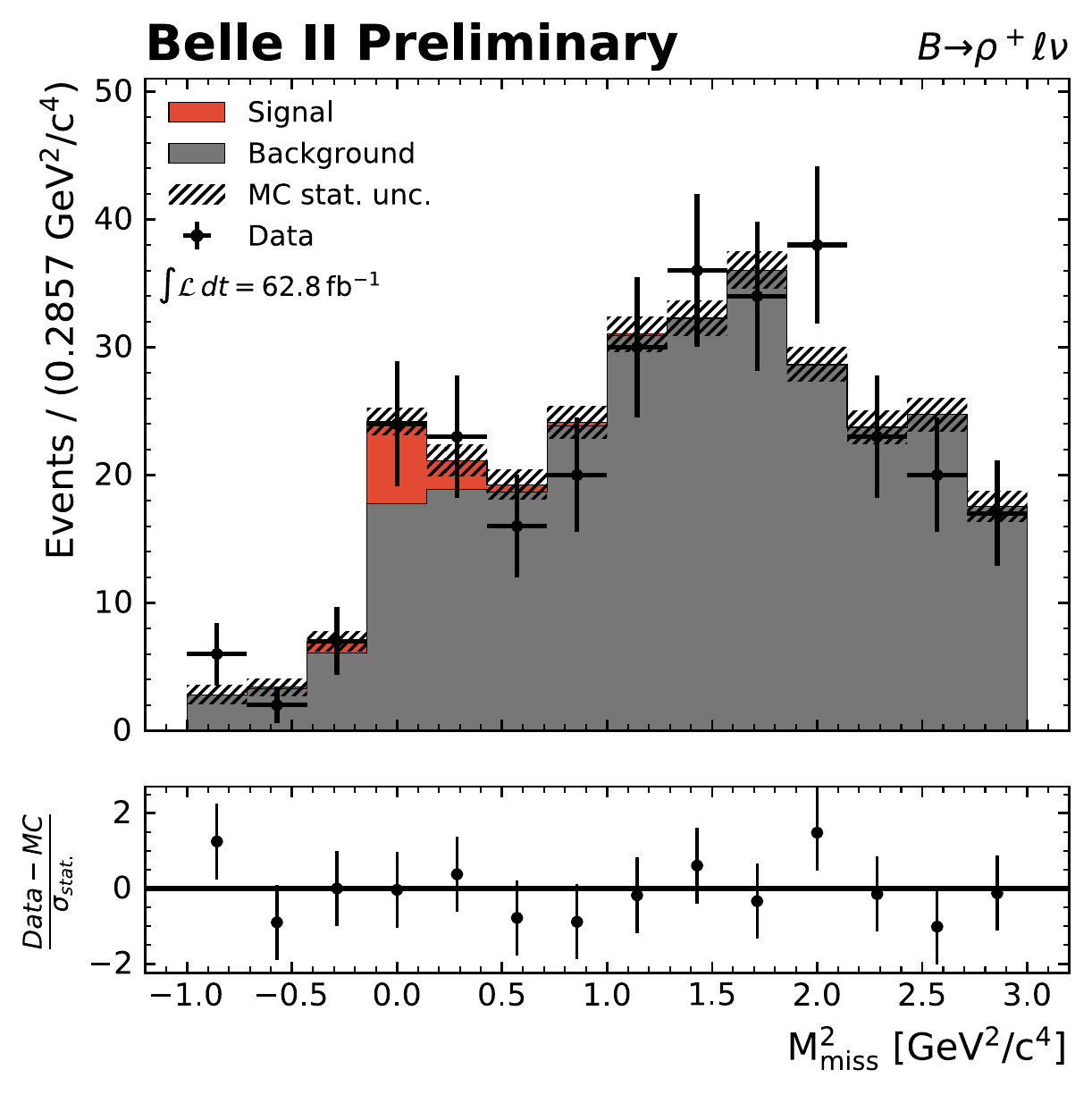}
\includegraphics[scale=0.6]{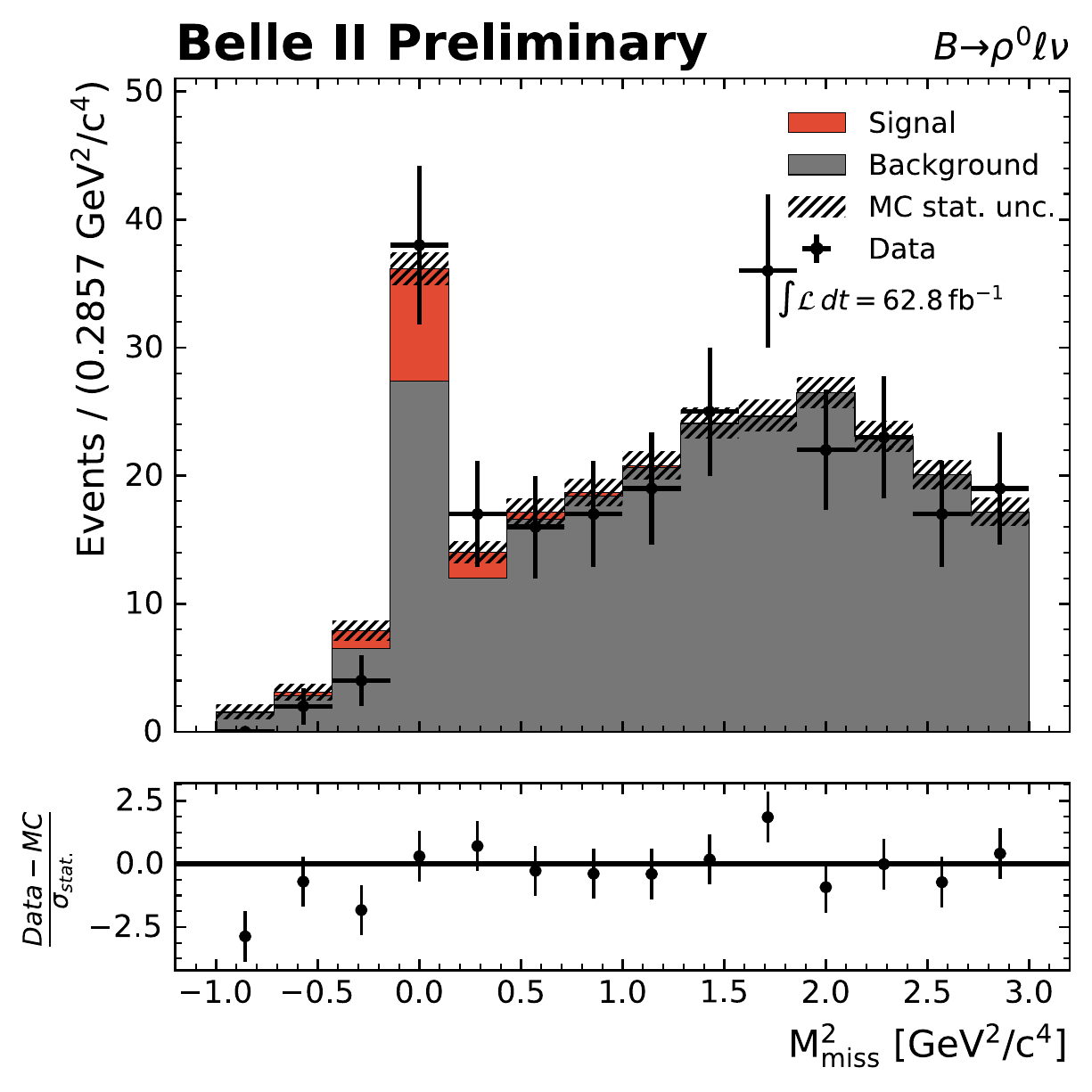}
    \caption{$M_{\mathrm{miss}}^2$ distributions for (left) $B^0 \to \rho^-\ell^+\nu_\ell$ and (right) $B^+ \to \rho^0\ell^+\nu_\ell$  candidates reconstructed from a sample corresponding to 62.8 \invfb of experimental data with fit projections overlaid.}
    \label{fig:postfitrholnu}
\end{center}
\end{figure}

Template probability density functions (PDFs) are subsequently built from the MC signal and background $M_{\mathrm{miss}}^2$ distributions shown in Figures \ref{fig:prefitpilnu} and \ref{fig:prefitrholnu}, normalized to the luminosity of the data sample. For each decay mode studied, due to limited sample size, all background components are combined together into a single background PDF, and a two-component extended unbinned maximum likelihood fit to the $M_{\mathrm{miss}}^2$ distributions in data under the signal plus background hypothesis is performed. The fit returns two parameters, namely the signal and background yields, which are allowed to float during the fit with no additional constraints. The resultant fitted distributions are illustrated in Figures \ref{fig:postfitpilnu} and \ref{fig:postfitrholnu} for $B\to\pi\ell\nu_\ell$ and $B\to\rho\ell\nu_\ell$ decays, respectively.

Fairly good agreement between simulated and measured data is observed across the $M_{\mathrm{miss}}^2$ distributions shown, including in the signal region. In the $B \to \pi \ell \nu_\ell$ case, a clear signal peak can be seen at $M_{\mathrm{miss}}^2$ = 0 for both data and MC, with all other backgrounds peaking at higher values of $M_{\mathrm{miss}}^2$. For $B \to \rho \ell \nu_\ell$ candidates, both the signal and background components produce a peak at $M_{\mathrm{miss}}^2$ = 0, with the background peak arising from the non-resonant component in MC. Here, all final-state particles excluding the neutrino in the $B \to \pi \pi \ell \nu_\ell$ decays are reconstructed and thus a peak at $M_{\mathrm{miss}}^2$ = 0 is expected. Using $M_{\mathrm{miss}}^2$, this background can therefore not be estimated directly and must be estimated based on its $\pi\pi$ mass distribution.

Additional extended maximum likelihood fits are then performed to the same data samples under the background-only hypothesis. The likelihood ratio $\lambda$ between both fits is computed for each decay mode, $\lambda = \mathcal{L}_{S+B}/\mathcal{L}_{B} \hspace{0.5em},$ where $\mathcal{L}_{S+B}$ and $\mathcal{L}_{B}$ are the maximised likelihoods returned by the fits to the background-only and signal + background hypotheses, respectively. A significance estimator $\Sigma$ is subsequently defined based on the likelihood ratio, $\Sigma = \sqrt{2\ln\lambda}\hspace{0.5em}$. The fitted yields for each decay mode are listed in Table \ref{table:datafitresults}, together with the observed significance.

\begin{table}
\caption{Yields obtained from the extended maximum likelihood fits to 62.8\invfb of data. The observed significances are also listed.}\label{table:datafitresults}
\def\arraystretch{1.2}%
\begin{tabular}{cccc}
\hline\hline
Process & Fitted signal yield & Fitted background yield & Observed significance \\
\hline
$B^0 \to \pi^- \ell^+ \nu_\ell$ & 37.1 $\pm$ 7.4 & 256.9 $\pm$ 16.6 & 7.7$\sigma$\\ 
$B^+ \to \pi^0 \ell^+ \nu_\ell$ & 34.8 $\pm$ 7.9 & 222.1 $\pm$ 15.8 & 6.2$\sigma$\\ 
$B^0 \to \rho^- \ell \nu$ &11.0 $\pm$ 8.3    & 331.0 $\pm$ 21.5    & 1.4$\sigma$\\ 
$B^+ \to \rho^0 \ell \nu$ & 13.7 $\pm$ 9.4    & 289.1 $\pm$ 21.2      & 1.5$\sigma$\\
\hline\hline
\end{tabular}
\end{table}

Due to the comparatively low reconstruction efficiency of the $B \to \rho \ell \nu_\ell$ decay modes, these do not reach the statistical threshold of evidence. However, as these are already known decay modes, we quote central values with uncertainties together with upper limits. The branching fractions for each decay mode are extracted using the following formulas:\\

\begin{equation}
\mathcal{B}(B^0 \to \pi^- \ell^+ \nu_\ell, B^0 \to \rho^- \ell^+ \nu_\ell) = \frac{N_{\mathrm{sig}}^{\mathrm{data}}(1 + f_{\mathrm{+0}})}{4\times \mathrm{CF}_{\mathrm{FEI}} (\times  \mathrm{SF}_{\pi^0})  \times N_{B\bar{B}} \times \epsilon} \hspace{0.5em},
\end{equation}
\begin{equation}
\mathcal{B}(B^+ \to \pi^0 \ell^+ \nu_\ell, B^+ \to \rho^0 \ell^+ \nu_\ell) = \frac{N_{\mathrm{sig}}^{\mathrm{data}}(1 + f_{\mathrm{+0}})}{4\times \mathrm{CF}_{\mathrm{FEI}} (\times  \mathrm{SF}_{\pi^0}) \times f_{\mathrm{+0}} \times N_{B\bar{B}} \times \epsilon} \hspace{0.5em},
\end{equation}
where $N_{\mathrm{sig}}^{\mathrm{data}}$ is the fitted signal yield obtained from data, $f_{\mathrm{+0}}$ is the ratio between the branching fractions of the decays of the $\Upsilon$(4S) meson to pairs of charged and neutral $B$-mesons \cite{Zyla:2020zbs}, $\mathrm{CF}_{\mathrm{FEI}}$ is the FEI calibration factor, $\mathrm{SF}_{\pi^0}$ is a scaling factor to correct the $\pi^0$ reconstruction efficiency (for $B^+ \to \pi^0 \ell^+ \nu_\ell$ and  $B^0 \to \rho^- \ell^+ \nu_\ell$ only), $N_{B\bar{B}}$ is the number of $B$-meson pairs counted in the current data set, and $\epsilon$ is the reconstruction efficiency. The factor of four present in the denominator accounts for the two $B$-mesons in the $\Upsilon$(4S) decay and the reconstruction of both light lepton flavors. The signal efficiency is calculated from the ratio of signal events present in MC before and after all analysis selections.

In the case of $B \to \rho \ell \nu_\ell$ decay modes, no significant signal is observed. Therefore, a 95\% confidence level (CL) upper limit on the branching fraction is calculated assuming a Gaussian likelihood for the branching fractions.
The values of the above parameters together with the measured branching fractions and upper limits are summarised in Tables \ref{table:BFpi} and \ref{table:BFrho} for $B\to\pi\ell\nu_\ell$ and $B\to\rho\ell\nu_\ell$ decays, respectively. The branching fractions agree well with the current world averages for both $B \to \pi \ell \nu_\ell$ decay modes \cite{Zyla:2020zbs}. All branching fraction uncertainties are largely dominated by sample size at the current integrated luminosity.

\begin{table} [h!]
\caption{Measured branching fractions of $B^0 \to \pi^- \ell^+ \nu_\ell$ and $B^+ \to \pi^0 \ell^+ \nu_\ell$ decays using 62.8\invfb of data, compared with the current world averages. The values of the parameters used in the measurement are also given.}\label{table:BFpi}
\def\arraystretch{1.2}%
\begin{tabular}{ccc}
\hline\hline
                                    & $B^0 \to \pi^- \ell^+ \nu_\ell$                        & $B^+ \to \pi^0 \ell^+ \nu_\ell$\\
                                    \hline
 $N_{\mathrm{sig}}^{\mathrm{data}}$ & 37.1 $\pm$ 7.4                                        & 34.8 $\pm$ 7.9\\   $f_{\mathrm{+0}}$                  & \multicolumn{2}{c}{1.058 $\pm$ 0.024}\\  
 $\mathrm{CF}_{\mathrm{FEI}}$       & 0.79 $\pm$ 0.02                                     & $0.63 \pm 0.02$\\  
 $\mathrm{SF}_{\pi^0}$              & -                                                       & $0.945 \pm $ 0.041\\  
 $N_{B\bar{B}}$                     & \multicolumn{2}{c}{(68.21 $\pm$ 0.75) $\times 10^6$} \\  
 $\epsilon$                         & (0.254 $\pm$ 0.001)$\%$                                 & (0.506 $\pm$ 0.002)$\%$ \\  
 $\mathcal{B}$                      & (1.40 $\pm$ 0.28(stat) $\pm$ 0.05(syst)) $\times 10^{-4}$ & (8.29 $\pm$ 1.99(stat) $\pm$ 0.46(syst)) $\times 10^{-5}$ \\  
 $\mathcal{B}_{\mathrm{PDG}}$       & (1.50 $\pm$ 0.06) $\times 10^{-4}$                                               & (7.80 $\pm$ 0.27) $\times 10^{-5}$\\
 \hline\hline
\end{tabular}
\end{table}

\begin{table}
\def\arraystretch{1.2}%
\caption{Measured branching fractions and upper limits of $B^0 \to \rho^- \ell^+ \nu_\ell$ and $B^+ \to \rho^0 \ell^+ \nu_\ell$ decays using \lint of data, compared with the current world averages. The values of the parameters used in the measurement are also given.}\label{table:BFrho}
\begin{tabular}{ccc}
 \hline\hline
                                    & $B^0 \to \rho^- \ell^+ \nu_\ell$                                                  & $B^+ \to \rho^0 \ell^+ \nu_\ell$\\ 
\hline
 $N_{\mathrm{sig}}^{\mathrm{data}}$ & 11.0 $\pm$ 8.3                                                                 & 13.7 $\pm$ 9.4\\  
 $f_{\mathrm{+0}}$                  & \multicolumn{2}{c}{1.058 $\pm$ 0.024}                                               \\  
 $\mathrm{CF}_{\mathrm{FEI}}$       & $0.79 \pm 0.02$                                                              & $0.63 \pm 0.02$\\  
  $\mathrm{SF}_{\pi^0}$             & $0.945 \pm 0.041$                                                  & --- \\  
 $N_{B\bar{B}}$                     & \multicolumn{2}{c}{(68.21 $\pm$ 0.75) $\times 10^6$}                                \\  
 $\epsilon$                         & (0.073 $\pm$ 0.002)$\%$                                                          & (0.168 $\pm$ 0.001)$\%$ \\  
 $\mathcal{B}$                      & (1.51 $\pm$ 1.13(stat) $\pm$ 0.09(syst)) $\times 10^{-4}$ & (9.26 $\pm$ 6.33(stat) $\pm$ 0.38(syst)) $\times 10^{-5}$ \\
 $95\%$ CL limit              & $< 3.37 \times 10^{-4}$                                     & $< 19.7 \times 10^{-5}$  \\
 $\mathcal{B}_{\mathrm{PDG}}$       & ($2.94 \pm 0.11 \pm 0.18$) $\times 10^{-4}$                                     & ($1.58 \pm 0.11$) $\times 10^{-4}$\\
 \hline\hline
\end{tabular}
\end{table}

\begin{figure}[h!]
\begin{center}
\includegraphics[scale=0.6]{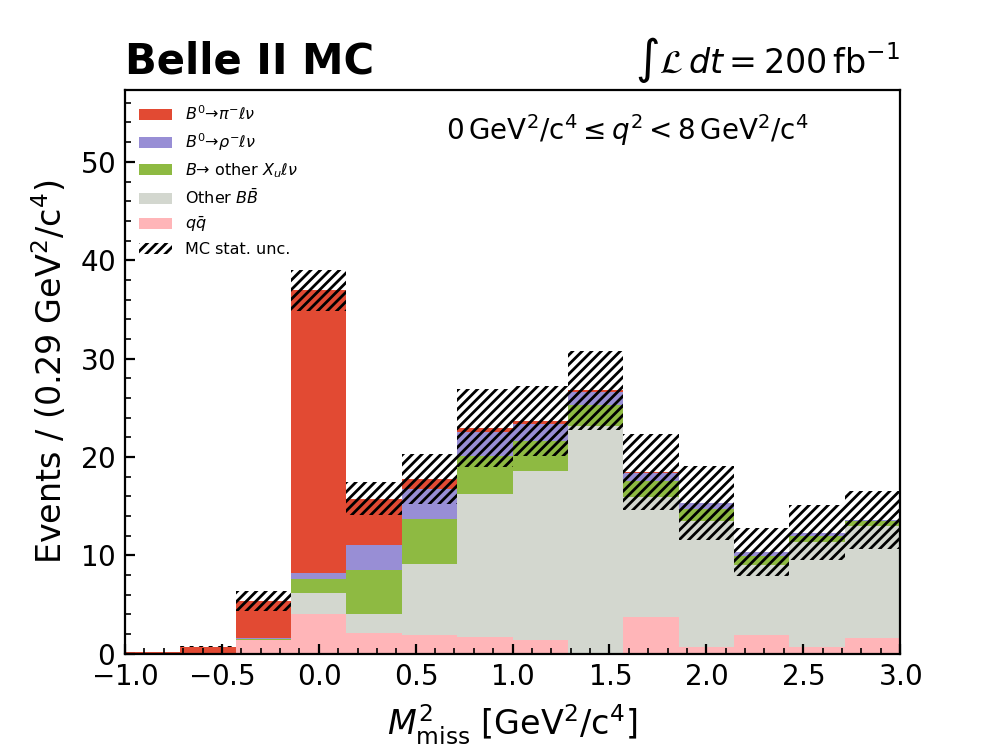}
\includegraphics[scale=0.36]{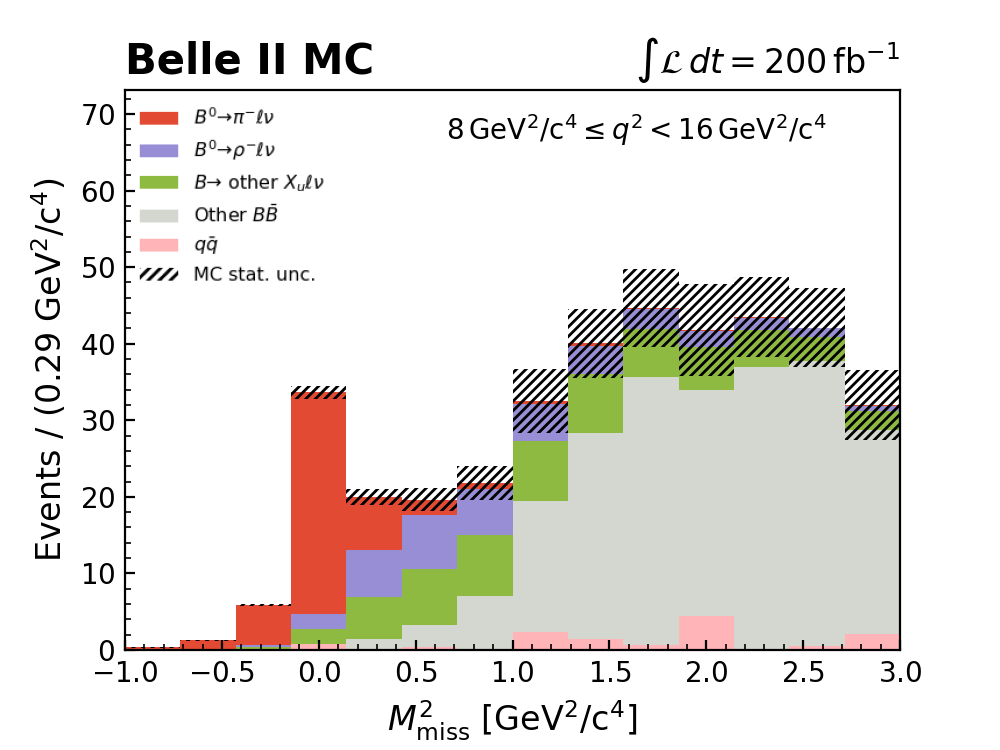}
\includegraphics[scale=0.36]{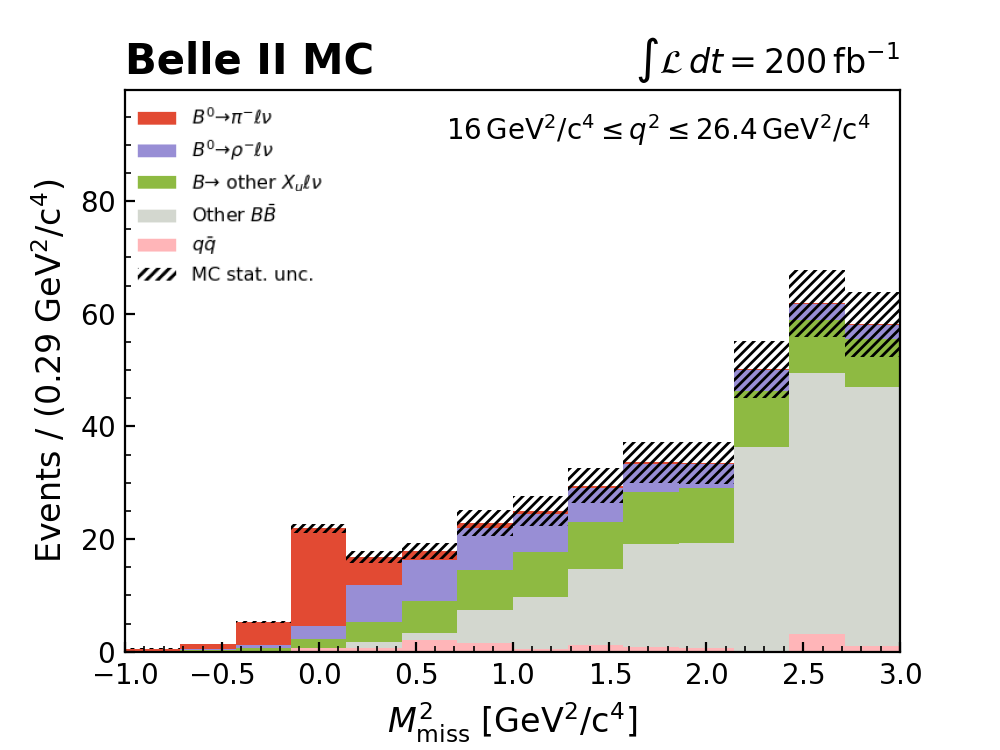}
  \caption{$M_{\mathrm{miss}}^2$ distributions for $B^0 \to \pi^-\ell^+\nu_\ell$ candidates restricted to three bins in $q^2$ and reconstructed from a sample corresponding to 200 \invfb of simulated data.}
  \label{fig:prefitpilnuq2}
\end{center}
\end{figure}

\begin{figure}[h!]
\begin{center}
\includegraphics[scale=0.6]{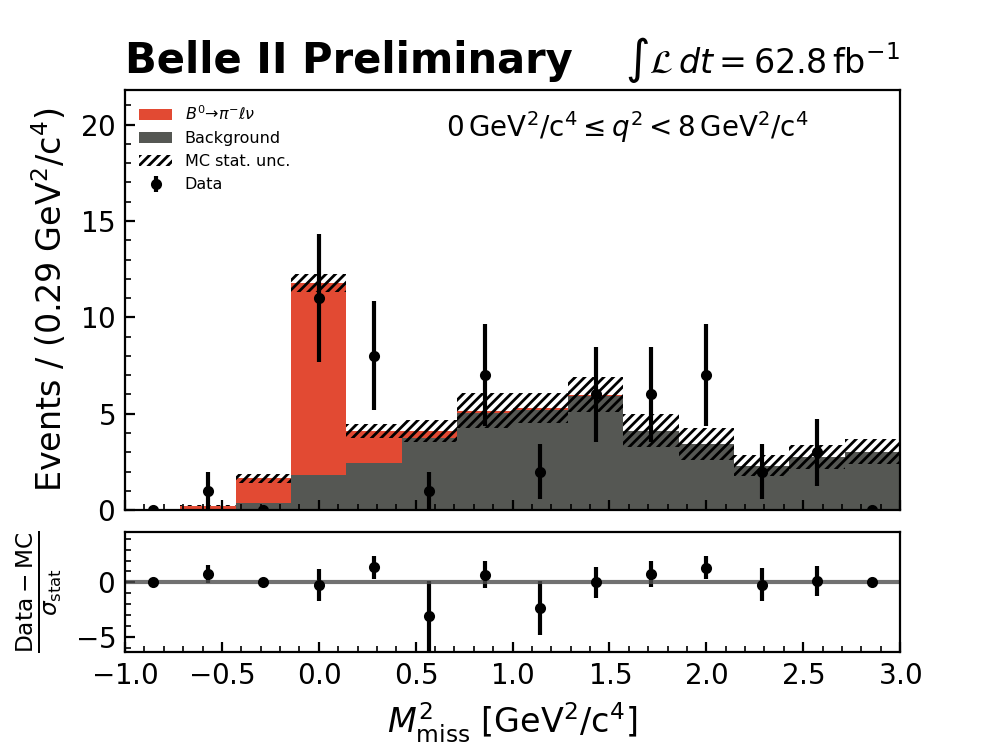}
\includegraphics[scale=0.36]{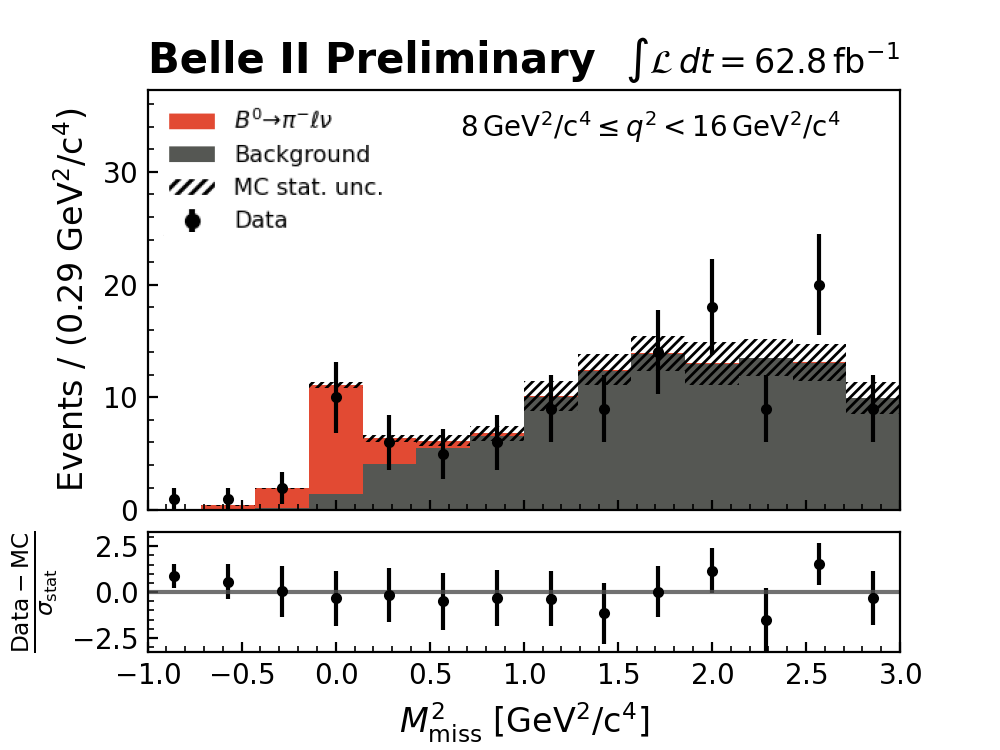}
\includegraphics[scale=0.36]{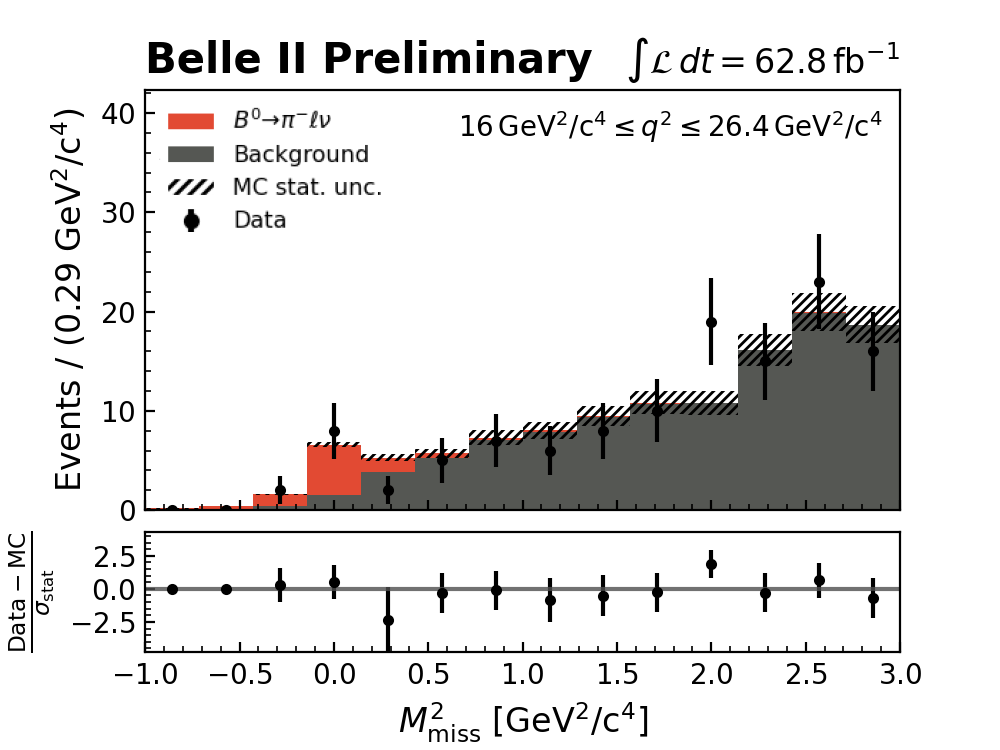}
  \caption{$M_{\mathrm{miss}}^2$ for $B^0 \to \pi^-\ell^+\nu_\ell$ decays candidates restricted to three bins in $q^2$ and reconstructed from a sample corresponding to 62.8 \invfb of simulated data  with fit projections overlaid.}
  \label{fig:postfitpilnuq2}
\end{center}
\end{figure}

In preparation for a future extraction of the Cabibbo-Kobayashi-Maskawa matrix-element magnitude $|V_{\mathrm{ub}}|$, fits to the $B^0\to\pi^-\ell^+\nu_\ell$ $M_{\mathrm{miss}}^2$ distribution are also performed in bins of the squared momentum transfer to the leptonic system, $q^2$. Due to the small sample size, only three bins are considered, with $0 \leq q^2 < 8$ $\text{GeV}^2/c^4 $, $8 \leq q^2 < 16$ $\text{GeV}^2/c^4$ and $16 \leq q^2 \leq 26.4$ $\text{GeV}^2/c^4$, respectively. The $B^+\to\pi^0\ell^+\nu_\ell$ mode is not yet considered for this treatment due to its lower significance and signal yields. Figures \ref{fig:prefitpilnuq2} and \ref{fig:postfitpilnuq2} display the $M_{\mathrm{miss}}^2$ distributions for the full MC sample used for the analysis, and the equivalent post-fit distributions in data for each of these $q^2$ bins. The data-MC agreement is fair across the three distributions, with clear signal peaks identified.

The partial branching fractions for each $q^2$ bin are then evaluated via the following formula:\\
\begin{equation}
\Delta \mathcal{B}_i(B^0 \to \pi^- \ell^+ \nu_\ell) = \frac{N_{\mathrm{sig, i}}^{\mathrm{data}}(1 + f_{\mathrm{+0}})}{4\times \mathrm{CF}_{\mathrm{FEI}} \times N_{B\bar{B}} \times \epsilon_i} \hspace{0.5em},
\end{equation}
which is equivalent to that for the total branching fraction with the exception that the signal efficiency $\epsilon_i$ and fitted yield $N_{\mathrm{sig}}^{\mathrm{data}}$ are now determined in each $q^2$ bin.

The fitted yields, observed significance and partial branching fractions for each $q^2$ bin are summarised in Table \ref{table:BFs_q2bins_B0pilnu}. The sum of the partial branching fractions is both consistent with and closer to the known value for $\mathcal{B}$($B^0\to\pi^-\ell^+\nu_\ell$) than the branching fraction obtained from the fit over the entire $q^2$ range.

\begin{table} [h!]
\footnotesize
\def\arraystretch{1}%
\caption{Fitted yields, observed significance and partial branching fractions obtained from fits to the $M_{\mathrm{miss}}^2$ distributions of $B^0\to\pi^-\ell^+\nu_\ell$ decays, in three bins of $q^2$. The signal efficiencies from MC are also listed.}\label{table:BFs_q2bins_B0pilnu}
\begin{tabular}{ccccc}
\hline\hline
$q^2$ bin &  Fitted yield  & MC efficiency $\epsilon_i$ & Significance & $\Delta \mathcal{B}$($B^0\to\pi^-\ell^+\nu_\ell$) \\\hline
{\footnotesize $0 \leq q^2 < 8$ GeV$^2/c^4$}  & 13.9 $\pm$ 4.5 & (0.230 $\pm$ 0.002)$\%$ & 4.9$\sigma$ & (0.58 $\pm$ 0.19(stat) $\pm$ 0.02(syst)) $\times 10^{-4}$\\
{\footnotesize $8  \leq q^2 < 16$ GeV$^2/c^4 $} & 15.5 $\pm$ 4.6 & (0.276 $\pm$ 0.002)$\%$ & 5.6$\sigma$ & (0.54 $\pm$ 0.16(stat) $\pm$ 0.02(syst)) $\times 10^{-4}$\\
{\footnotesize $16 \leq q^2 \leq 26.4$  GeV$^2/c^4$}  & 9.4 $\pm$ 3.8 & (0.260 $\pm$ 0.002)$\%$ &  3.7$\sigma$ & (0.35 $\pm$ 0.14(stat) $\pm$ 0.01(syst)) $\times 10^{-4}$\\
 
Sum & 38.8 $\pm$ 7.5 & - & - & (1.47 $\pm$ 0.29(stat) $\pm$ 0.05(syst)) $\times 10^{-4}$\\
Fit over full $q^2$ range & 37.1 $\pm$ 7.4 & (0.254 $\pm$ 0.001)$\%$ & 7.7$\sigma$ & (1.40 $\pm$ 0.28(stat) $\pm$ 0.05(syst)) $\times 10^{-4}$\\
$\mathcal{B}_{\mathrm{PDG}}$ & - & - & - &(1.50 $\pm$ 0.06) $\times 10^{-4}$\\\hline\hline
\end{tabular}
\end{table}

\section{Systematic Uncertainties}
\label{sec:systematics}

A number of sources of systematic uncertainty are identified for this analysis and evaluated for the branching fraction measurements. The relative uncertainties for each source, in percent, are summarised in Tables \ref{table:syserrors} and \ref{table:syserrors_B0_q2bins} for the total and partial branching fraction measurements, respectively, and include

\begin{itemize}
    \item \boldmath $f_{\mathrm{+0}}$: \unboldmath We combine the errors on the world averages for the branching fractions $\mathcal{B}(\Upsilon$(4S)$\to B^+B^-$) and $\mathcal{B}(\Upsilon$(4S) $\to B^0\bar{B}^0$) and calculate the relative uncertainty on the fraction $f_{\mathrm{+0}}$.
    \item \textbf{FEI calibration:} The uncertainty on the FEI calibration factors are determined from fits to the lepton momentum spectrum of $B\to X \ell \nu_\ell$ decays. Sources of uncertainty in this fit include uncertainties on both the branching fractions and form factors of the various semileptonic components of $B\to X \ell \nu_\ell$, the lepton ID efficiency and fake rate uncertainties, tracking uncertainties and statistical uncertainties in the MC template distribution. The relative uncertainty on the calibration factor forms the dominant source of systematic uncertainty for the $B^0 \to \pi^- \ell^+ \nu_\ell$ and $B^+ \to \rho^0 \ell^+ \nu_\ell$ analyses.
    \item \textbf{$\pi^0$ efficiency:} The relative uncertainty on the scaling factor to correct the $\pi^0$ efficiency in MC is derived via an independent study of $\eta \to 3\pi^0$ decays. It forms the dominant source of systematic uncertainty for $B^+ \to \pi^0 \ell^+ \nu_\ell$ and $B^0 \to \rho^- \ell^+ \nu_\ell$.
    \item \boldmath $N_{B\bar{B}}$: \unboldmath The uncertainty on the number of $B\bar{B}$ events in the present data set includes systematic effects due to uncertainties on the luminosity, beam energy spread and shift, tracking efficiency and the selection efficiency of $B\bar{B}$ events.
    \item \textbf{Reconstruction efficiency:} We represent the uncertainty on the reconstruction efficiency with a binomial error dependent on the size of the MC samples used for the analysis.  
    \item \textbf{Tracking:} We assign a constant systematic uncertainty of 0.69$\%$ for each charged particle. For decay modes with multiple tracks we assume the associated uncertainties to be completely correlated.
    \item \textbf{Lepton identification:} The lepton efficiencies and pion/kaon fake rates are evaluated in bins of the lepton momentum $p$ and polar angle $\theta$, each with statistical and systematic uncertainties. The effect of these uncertainties on the signal reconstruction efficiency is determined through generating 200 variations on the nominal correction weights via Gaussian smearing. The relative uncertainty is then taken from the spread on the values of the reconstruction efficiency over all variations. 
    \item \textbf{Pion identification:} The pion efficiencies and kaon fake rates are similarly evaluated in bins of the pion momentum $p$ and the polar angle $\theta$, as is done for the lepton identification corrections. The relative systematic uncertainty for the pionID corrections is likewise determined via evaluating the effect of Gaussian smearing on the signal reconstruction efficiency, using 200 variations on the nominal correction weights.
\end{itemize}

\begin{table}
\centering
\caption{Sources of systematic uncertainty quoted as a percentage of the measured branching fractions.}\label{table:syserrors}
\begin{tabular}{ccccc}
\hline\hline
Source                                & $\%$  of                     & $\%$  of       & $\%$ of         & $\%$  of   \\
~                                     & \bfpic{}                     & \bfpiz{}       & \bfrhoc{}       & \bfrhoz{} \\
\hline
  
 FEI calibration                      & 2.8                         & 2.5           & 2.8            & 2.5 \\                   
 $N_{B\bar{B}}$                       & \multicolumn{4}{c}{1.1}\\  
 $f_{\mathrm{+0}}$                    & \multicolumn{4}{c}{1.2} \\  
 Reconstruction efficiency $\epsilon$ & 0.5                         & 0.5           & 0.6            & 0.6 \\                    
 Tracking                             & 1.4                         & 0.7           & 1.4            & 2.1 \\                    
 Lepton ID                            & 1.5                         & 1.5           & 1.1            & 1.1 \\                    
 Pion ID                              & 0.6                         & -              & 0.8            & 1.5 \\                    
 $\pi^0$ efficiency                   & -                            & 4.4           & 4.4            & --- \\           
 \hline
 Total                                & 3.9                         & 5.6           & 5.8            & 4.1 \\
 \hline\hline
\end{tabular}
\end{table}

For  $B \to \pi \ell \nu_\ell$ decays, the systematic uncertainties from the modeling of  $B \to X_u \ell \nu_\ell$ are expected to be small compared to other systematic uncertainties. For $B \to \rho \ell \nu_\ell$ decays, the uncertainty on the non-resonant model cannot be quantified with the currently available data set, but is expected to be small compared to the statistical uncertainties. Additional systematic uncertainties on the efficiencies of various selection criteria are not included, as these are expected to be considerably small in comparison to other systematic effects.

\begin{table} [h!]
\small
\caption{Sources of systematic uncertainty quoted as a percentage of the measured $B^0\to\pi^-\ell^+\nu_\ell$ partial branching fractions in three bins of $q^2$.}\label{table:syserrors_B0_q2bins}
\begin{tabular}{cccc}
\hline\hline
Source &   \multicolumn{3}{c}{$\%$ of $\Delta \mathcal{B}_i$($B^0\to\pi^-\ell^+\nu_\ell$)}\\
& {\footnotesize $0\leq q^2<8 \text{GeV}^2/c^4$} & {\footnotesize$8 \leq q^2<16 \text{GeV}^2/c^4 $} & {\footnotesize $16 \leq q^2\leq 26.4  \text{GeV}^2/c^4$}\\\hline
 $f_{\mathrm{+0}}$ & \multicolumn{3}{c}{1.2}\\
 FEI calibration &  \multicolumn{3}{c}{2.8} \\
 $N_{B\bar{B}}$ &  \multicolumn{3}{c}{1.1}\\
 Tracking & \multicolumn{3}{c}{1.4}\\
 Recon. efficiency $\epsilon_i$ & 0.8  & 0.8 & 0.9\\
 Lepton ID & 1.7 & 1.3 & 1.6\\
 Pion ID & 0.7 & 0.6 & 0.6\\\hline
 Total & 4.0  & 3.9  &  4.0\\\hline\hline
\end{tabular}

\end{table}

\section{Summary}
\label{sec:conclusions}
In summary, we present an analysis of the semileptonic decay $B^+ \to \pi^0 \ell^+ \nu_\ell$ via hadronic tagging in a Belle II data sample corresponding to 62.8\invfb. A branching fraction of (8.29 $\pm$ 1.99(stat) $\pm$ 0.46(syst)) $\times 10^{-5}$ is measured with an observed signal significance of 6.2$\sigma$, in agreement with the current world average \cite{Zyla:2020zbs}. We also present results for $B^+ \to \rho^0 \ell^+ \nu_\ell$ and $B^0 \to \rho^- \ell^+ \nu_\ell$ with an observed significance of 1.5$\sigma$ and 1.4$\sigma$, respectively. The branching fractions are measured for each decay, with \bfrhoz{} = (9.26 $\pm$ 6.33(stat) $\pm$ 0.38(syst)) $\times 10^{-5}$ and \bfrhoc = (1.51 $\pm$ 1.13(stat) $\pm$ 0.09(syst)) $\times 10^{-4}$, corresponding to $95\%$ CL upper limits with \bfrhoz{} $ < 19.7 \times 10^{-5}$ and \bfrhoc{} $ < 3.37 \times 10^{-4}$. For $B^+ \to \rho^0 \ell^+ \nu_\ell$, the branching fraction is found to be in agreement with the current world average \cite{Zyla:2020zbs}. An updated branching fraction for the $B^0 \to \pi^- \ell^+ \nu_\ell$ decay, (1.47 $\pm$ 0.29(stat) $\pm$ 0.05(syst)) $\times 10^{-4}$, is quoted based on the sum of the partial branching fractions in three bins of the momentum transfer to the leptonic system, $q^2$.

\section*{Acknowledgements}
\label{sec:acknowledgements}
\input{acknowledgements.tex}

\bibliography{belle2}
\bibliographystyle{belle2-note}

\end{document}

%% file: authors-conf2021.tex
\newcommand{\instCPPM}{Aix Marseille Universit\'{e}, CNRS/IN2P3, CPPM, 13288 Marseille, France}
\newcommand{\instYerevan}{Alikhanyan National Science Laboratory, Yerevan 0036, Armenia}
\newcommand{\instBeihang}{Beihang University, Beijing 100191, China}
\newcommand{\instBNL}{Brookhaven National Laboratory, Upton, New York 11973, U.S.A.}
\newcommand{\instBINP}{Budker Institute of Nuclear Physics SB RAS, Novosibirsk 630090, Russian Federation}
\newcommand{\instCMU}{Carnegie Mellon University, Pittsburgh, Pennsylvania 15213, U.S.A.}
\newcommand{\instCinvestavIPN}{Centro de Investigacion y de Estudios Avanzados del Instituto Politecnico Nacional, Mexico City 07360, Mexico}
\newcommand{\instPrague}{Faculty of Mathematics and Physics, Charles University, 121 16 Prague, Czech Republic}
\newcommand{\instChiangMai}{Chiang Mai University, Chiang Mai 50202, Thailand}
\newcommand{\instChiba}{Chiba University, Chiba 263-8522, Japan}
\newcommand{\instChonnam}{Chonnam National University, Gwangju 61186, South Korea}
\newcommand{\instConacyt}{Consejo Nacional de Ciencia y Tecnolog\'{\i}a, Mexico City 03940, Mexico}
\newcommand{\instDESY}{Deutsches Elektronen--Synchrotron, 22607 Hamburg, Germany}
\newcommand{\instDuke}{Duke University, Durham, North Carolina 27708, U.S.A.}
\newcommand{\instITAR}{Institute of Theoretical and Applied Research (ITAR), Duy Tan University, Hanoi 100000, Vietnam}
\newcommand{\instRomaENEA}{ENEA Casaccia, I-00123 Roma, Italy}
\newcommand{\instFuJen}{Department of Physics, Fu Jen Catholic University, Taipei 24205, Taiwan}
\newcommand{\instFudan}{Key Laboratory of Nuclear Physics and Ion-beam Application (MOE) and Institute of Modern Physics, Fudan University, Shanghai 200443, China}
\newcommand{\instGoettingen}{II. Physikalisches Institut, Georg-August-Universit\"{a}t G\"{o}ttingen, 37073 G\"{o}ttingen, Germany}
\newcommand{\instGifu}{Gifu University, Gifu 501-1193, Japan}
\newcommand{\instSOKENDAI}{The Graduate University for Advanced Studies (SOKENDAI), Hayama 240-0193, Japan}
\newcommand{\instGyeongsang}{Gyeongsang National University, Jinju 52828, South Korea}
\newcommand{\instHanyang}{Department of Physics and Institute of Natural Sciences, Hanyang University, Seoul 04763, South Korea}
\newcommand{\instKEK}{High Energy Accelerator Research Organization (KEK), Tsukuba 305-0801, Japan}
\newcommand{\instJPARC}{J-PARC Branch, KEK Theory Center, High Energy Accelerator Research Organization (KEK), Tsukuba 305-0801, Japan}
\newcommand{\instHiroshima}{Hiroshima University, Higashi-Hiroshima, Hiroshima 739-8530, Japan}
\newcommand{\instFrascati}{INFN Laboratori Nazionali di Frascati, I-00044 Frascati, Italy}
\newcommand{\instNapoliINFN}{INFN Sezione di Napoli, I-80126 Napoli, Italy}
\newcommand{\instPadovaINFN}{INFN Sezione di Padova, I-35131 Padova, Italy}
\newcommand{\instPerugiaINFN}{INFN Sezione di Perugia, I-06123 Perugia, Italy}
\newcommand{\instPisaINFN}{INFN Sezione di Pisa, I-56127 Pisa, Italy}
\newcommand{\instRomaINFN}{INFN Sezione di Roma, I-00185 Roma, Italy}
\newcommand{\instRomaTreINFN}{INFN Sezione di Roma Tre, I-00146 Roma, Italy}
\newcommand{\instTorinoINFN}{INFN Sezione di Torino, I-10125 Torino, Italy}
\newcommand{\instTriesteINFN}{INFN Sezione di Trieste, I-34127 Trieste, Italy}
\newcommand{\instIISER}{Indian Institute of Science Education and Research Mohali, SAS Nagar, 140306, India}
\newcommand{\instIITBhubaneswar}{Indian Institute of Technology Bhubaneswar, Satya Nagar 751007, India}
\newcommand{\instIITGuwahati}{Indian Institute of Technology Guwahati, Assam 781039, India}
\newcommand{\instIITHyderabad}{Indian Institute of Technology Hyderabad, Telangana 502285, India}
\newcommand{\instIITMadras}{Indian Institute of Technology Madras, Chennai 600036, India}
\newcommand{\instIndiana}{Indiana University, Bloomington, Indiana 47408, U.S.A.}
\newcommand{\instIHEPRussia}{Institute for High Energy Physics, Protvino 142281, Russian Federation}
\newcommand{\instHEPHYVienna}{Institute of High Energy Physics, Vienna 1050, Austria}
\newcommand{\instIHEPChina}{Institute of High Energy Physics, Chinese Academy of Sciences, Beijing 100049, China}
\newcommand{\instIPP}{Institute of Particle Physics (Canada), Victoria, British Columbia V8W 2Y2, Canada}
\newcommand{\instIOP}{Institute of Physics, Vietnam Academy of Science and Technology (VAST), Hanoi, Vietnam}
\newcommand{\instIFIC}{Instituto de Fisica Corpuscular, Paterna 46980, Spain}
\newcommand{\instISU}{Iowa State University, Ames, Iowa 50011, U.S.A.}
\newcommand{\instJAEA}{Advanced Science Research Center, Japan Atomic Energy Agency, Naka 319-1195, Japan}
\newcommand{\instMainz}{Institut f\"{u}r Kernphysik, Johannes Gutenberg-Universit\"{a}t Mainz, D-55099 Mainz, Germany}
\newcommand{\instGiessen}{Justus-Liebig-Universit\"{a}t Gie\ss{}en, 35392 Gie\ss{}en, Germany}
\newcommand{\instKarlsruhe}{Institut f\"{u}r Experimentelle Teilchenphysik, Karlsruher Institut f\"{u}r Technologie, 76131 Karlsruhe, Germany}
\newcommand{\instKitasato}{Kitasato University, Sagamihara 252-0373, Japan}
\newcommand{\instKISTI}{Korea Institute of Science and Technology Information, Daejeon 34141, South Korea}
\newcommand{\instKoreaUnivKU}{Korea University, Seoul 02841, South Korea}
\newcommand{\instKSU}{Kyoto Sangyo University, Kyoto 603-8555, Japan}
\newcommand{\instKyungpook}{Kyungpook National University, Daegu 41566, South Korea}
\newcommand{\instLPI}{P.N. Lebedev Physical Institute of the Russian Academy of Sciences, Moscow 119991, Russian Federation}
\newcommand{\instLNNU}{Liaoning Normal University, Dalian 116029, China}
\newcommand{\instLMU}{Ludwig Maximilians University, 80539 Munich, Germany}
\newcommand{\instLuther}{Luther College, Decorah, Iowa 52101, U.S.A.}
\newcommand{\instMNITJaipur}{Malaviya National Institute of Technology Jaipur, Jaipur 302017, India}
\newcommand{\instMPP}{Max-Planck-Institut f\"{u}r Physik, 80805 M\"{u}nchen, Germany}
\newcommand{\instMPGHLL}{Semiconductor Laboratory of the Max Planck Society, 81739 M\"{u}nchen, Germany}
\newcommand{\instMcGill}{McGill University, Montr\'{e}al, Qu\'{e}bec, H3A 2T8, Canada}
\newcommand{\instMEPhI}{Moscow Physical Engineering Institute, Moscow 115409, Russian Federation}
\newcommand{\instNagoya}{Graduate School of Science, Nagoya University, Nagoya 464-8602, Japan}
\newcommand{\instNagoyaIAR}{Institute for Advanced Research, Nagoya University, Nagoya 464-8602, Japan}
\newcommand{\instNagoyaKMI}{Kobayashi-Maskawa Institute, Nagoya University, Nagoya 464-8602, Japan}
\newcommand{\instNaraWu}{Nara Women's University, Nara 630-8506, Japan}
\newcommand{\instHSE}{National Research University Higher School of Economics, Moscow 101000, Russian Federation}
\newcommand{\instNTUTaiwan}{Department of Physics, National Taiwan University, Taipei 10617, Taiwan}
\newcommand{\instNUUTaiwan}{National United University, Miao Li 36003, Taiwan}
\newcommand{\instKrakow}{H. Niewodniczanski Institute of Nuclear Physics, Krakow 31-342, Poland}
\newcommand{\instNiigata}{Niigata University, Niigata 950-2181, Japan}
\newcommand{\instNSU}{Novosibirsk State University, Novosibirsk 630090, Russian Federation}
\newcommand{\instOkinawa}{Okinawa Institute of Science and Technology, Okinawa 904-0495, Japan}
\newcommand{\instOsakaCity}{Osaka City University, Osaka 558-8585, Japan}
\newcommand{\instRCNP}{Research Center for Nuclear Physics, Osaka University, Osaka 567-0047, Japan}
\newcommand{\instPNNL}{Pacific Northwest National Laboratory, Richland, Washington 99352, U.S.A.}
\newcommand{\instPanjab}{Panjab University, Chandigarh 160014, India}
\newcommand{\instPanjabPAU}{Punjab Agricultural University, Ludhiana 141004, India}
\newcommand{\instRIKENMSL}{Meson Science Laboratory, Cluster for Pioneering Research, RIKEN, Saitama 351-0198, Japan}
\newcommand{\instXavier}{St. Francis Xavier University, Antigonish, Nova Scotia, B2G 2W5, Canada}
\newcommand{\instSeoul}{Seoul National University, Seoul 08826, South Korea}
\newcommand{\instSPU}{Showa Pharmaceutical University, Tokyo 194-8543, Japan}
\newcommand{\instSoochow}{Soochow University, Suzhou 215006, China}
\newcommand{\instSoongsil}{Soongsil University, Seoul 06978, South Korea}
\newcommand{\instLjubljanaJSI}{J. Stefan Institute, 1000 Ljubljana, Slovenia}
\newcommand{\instKyiv}{Taras Shevchenko National Univ. of Kiev, Kiev, Ukraine}
\newcommand{\instTata}{Tata Institute of Fundamental Research, Mumbai 400005, India}
\newcommand{\instTUM}{Department of Physics, Technische Universit\"{a}t M\"{u}nchen, 85748 Garching, Germany}
\newcommand{\instTelAviv}{Tel Aviv University, School of Physics and Astronomy, Tel Aviv, 69978, Israel}
\newcommand{\instToho}{Toho University, Funabashi 274-8510, Japan}
\newcommand{\instTohoku}{Department of Physics, Tohoku University, Sendai 980-8578, Japan}
\newcommand{\instTitech}{Tokyo Institute of Technology, Tokyo 152-8550, Japan}
\newcommand{\instTokyoMetropolitan}{Tokyo Metropolitan University, Tokyo 192-0397, Japan}
\newcommand{\instUAS}{Universidad Autonoma de Sinaloa, Sinaloa 80000, Mexico}
\newcommand{\instNapoliUNIV}{Dipartimento di Scienze Fisiche, Universit\`{a} di Napoli Federico II, I-80126 Napoli, Italy}
\newcommand{\instPadovaUNIV}{Dipartimento di Fisica e Astronomia, Universit\`{a} di Padova, I-35131 Padova, Italy}
\newcommand{\instPerugiaUNIV}{Dipartimento di Fisica, Universit\`{a} di Perugia, I-06123 Perugia, Italy}
\newcommand{\instPisaUNIV}{Dipartimento di Fisica, Universit\`{a} di Pisa, I-56127 Pisa, Italy}
\newcommand{\instRomaTreUNIV}{Dipartimento di Matematica e Fisica, Universit\`{a} di Roma Tre, I-00146 Roma, Italy}
\newcommand{\instTorinoUNIV}{Dipartimento di Fisica, Universit\`{a} di Torino, I-10125 Torino, Italy}
\newcommand{\instTriesteUNIV}{Dipartimento di Fisica, Universit\`{a} di Trieste, I-34127 Trieste, Italy}
\newcommand{\instMontreal}{Universit\'{e} de Montr\'{e}al, Physique des Particules, Montr\'{e}al, Qu\'{e}bec, H3C 3J7, Canada}
\newcommand{\instIJCLab}{Universit\'{e} Paris-Saclay, CNRS/IN2P3, IJCLab, 91405 Orsay, France}
\newcommand{\instIPHC}{Universit\'{e} de Strasbourg, CNRS, IPHC, UMR 7178, 67037 Strasbourg, France}
\newcommand{\instAdelaide}{Department of Physics, University of Adelaide, Adelaide, South Australia 5005, Australia}
\newcommand{\instBonn}{University of Bonn, 53115 Bonn, Germany}
\newcommand{\instUBC}{University of British Columbia, Vancouver, British Columbia, V6T 1Z1, Canada}
\newcommand{\instCincinnati}{University of Cincinnati, Cincinnati, Ohio 45221, U.S.A.}
\newcommand{\instFlorida}{University of Florida, Gainesville, Florida 32611, U.S.A.}
\newcommand{\instHawaii}{University of Hawaii, Honolulu, Hawaii 96822, U.S.A.}
\newcommand{\instHeidelberg}{University of Heidelberg, 68131 Mannheim, Germany}
\newcommand{\instLjubljanaUniLJ}{Faculty of Mathematics and Physics, University of Ljubljana, 1000 Ljubljana, Slovenia}
\newcommand{\instLouisville}{University of Louisville, Louisville, Kentucky 40292, U.S.A.}
\newcommand{\instMalaya}{National Centre for Particle Physics, University Malaya, 50603 Kuala Lumpur, Malaysia}
\newcommand{\instLjubljanaUM}{Faculty of Chemistry and Chemical Engineering, University of Maribor, 2000 Maribor, Slovenia}
\newcommand{\instMelbourne}{School of Physics, University of Melbourne, Victoria 3010, Australia}
\newcommand{\instMississippi}{University of Mississippi, University, Mississippi 38677, U.S.A.}
\newcommand{\instUOM}{University of Miyazaki, Miyazaki 889-2192, Japan}
\newcommand{\instPittsburgh}{University of Pittsburgh, Pittsburgh, Pennsylvania 15260, U.S.A.}
\newcommand{\instUSTC}{University of Science and Technology of China, Hefei 230026, China}
\newcommand{\instSAlabama}{University of South Alabama, Mobile, Alabama 36688, U.S.A.}
\newcommand{\instSCarolina}{University of South Carolina, Columbia, South Carolina 29208, U.S.A.}
\newcommand{\instSydney}{School of Physics, University of Sydney, New South Wales 2006, Australia}
\newcommand{\instUTokyo}{Department of Physics, University of Tokyo, Tokyo 113-0033, Japan}
\newcommand{\instEri}{Earthquake Research Institute, University of Tokyo, Tokyo 113-0032, Japan}
\newcommand{\instIPMU}{Kavli Institute for the Physics and Mathematics of the Universe (WPI), University of Tokyo, Kashiwa 277-8583, Japan}
\newcommand{\instVictoria}{University of Victoria, Victoria, British Columbia, V8W 3P6, Canada}
\newcommand{\instVPI}{Virginia Polytechnic Institute and State University, Blacksburg, Virginia 24061, U.S.A.}
\newcommand{\instWayneState}{Wayne State University, Detroit, Michigan 48202, U.S.A.}
\newcommand{\instYamagata}{Yamagata University, Yamagata 990-8560, Japan}
\newcommand{\instYonsei}{Yonsei University, Seoul 03722, South Korea}
\affiliation{\instCPPM}
\affiliation{\instYerevan}
\affiliation{\instBeihang}
\affiliation{\instBNL}
\affiliation{\instBINP}
\affiliation{\instCMU}
\affiliation{\instCinvestavIPN}
\affiliation{\instPrague}
\affiliation{\instChiangMai}
\affiliation{\instChiba}
\affiliation{\instChonnam}
\affiliation{\instConacyt}
\affiliation{\instDESY}
\affiliation{\instDuke}
\affiliation{\instITAR}
\affiliation{\instRomaENEA}
\affiliation{\instFuJen}
\affiliation{\instFudan}
\affiliation{\instGoettingen}
\affiliation{\instGifu}
\affiliation{\instSOKENDAI}
\affiliation{\instGyeongsang}
\affiliation{\instHanyang}
\affiliation{\instKEK}
\affiliation{\instJPARC}
\affiliation{\instHiroshima}
\affiliation{\instFrascati}
\affiliation{\instNapoliINFN}
\affiliation{\instPadovaINFN}
\affiliation{\instPerugiaINFN}
\affiliation{\instPisaINFN}
\affiliation{\instRomaINFN}
\affiliation{\instRomaTreINFN}
\affiliation{\instTorinoINFN}
\affiliation{\instTriesteINFN}
\affiliation{\instIISER}
\affiliation{\instIITBhubaneswar}
\affiliation{\instIITGuwahati}
\affiliation{\instIITHyderabad}
\affiliation{\instIITMadras}
\affiliation{\instIndiana}
\affiliation{\instIHEPRussia}
\affiliation{\instHEPHYVienna}
\affiliation{\instIHEPChina}
\affiliation{\instIPP}
\affiliation{\instIOP}
\affiliation{\instIFIC}
\affiliation{\instISU}
\affiliation{\instJAEA}
\affiliation{\instMainz}
\affiliation{\instGiessen}
\affiliation{\instKarlsruhe}
\affiliation{\instKitasato}
\affiliation{\instKISTI}
\affiliation{\instKoreaUnivKU}
\affiliation{\instKSU}
\affiliation{\instKyungpook}
\affiliation{\instLPI}
\affiliation{\instLNNU}
\affiliation{\instLMU}
\affiliation{\instLuther}
\affiliation{\instMNITJaipur}
\affiliation{\instMPP}
\affiliation{\instMPGHLL}
\affiliation{\instMcGill}
\affiliation{\instMEPhI}
\affiliation{\instNagoya}
\affiliation{\instNagoyaIAR}
\affiliation{\instNagoyaKMI}
\affiliation{\instNaraWu}
\affiliation{\instHSE}
\affiliation{\instNTUTaiwan}
\affiliation{\instNUUTaiwan}
\affiliation{\instKrakow}
\affiliation{\instNiigata}
\affiliation{\instNSU}
\affiliation{\instOkinawa}
\affiliation{\instOsakaCity}
\affiliation{\instRCNP}
\affiliation{\instPNNL}
\affiliation{\instPanjab}
\affiliation{\instPanjabPAU}
\affiliation{\instRIKENMSL}
\affiliation{\instXavier}
\affiliation{\instSeoul}
\affiliation{\instSPU}
\affiliation{\instSoochow}
\affiliation{\instSoongsil}
\affiliation{\instLjubljanaJSI}
\affiliation{\instKyiv}
\affiliation{\instTata}
\affiliation{\instTUM}
\affiliation{\instTelAviv}
\affiliation{\instToho}
\affiliation{\instTohoku}
\affiliation{\instTitech}
\affiliation{\instTokyoMetropolitan}
\affiliation{\instUAS}
\affiliation{\instNapoliUNIV}
\affiliation{\instPadovaUNIV}
\affiliation{\instPerugiaUNIV}
\affiliation{\instPisaUNIV}
\affiliation{\instRomaTreUNIV}
\affiliation{\instTorinoUNIV}
\affiliation{\instTriesteUNIV}
\affiliation{\instMontreal}
\affiliation{\instIJCLab}
\affiliation{\instIPHC}
\affiliation{\instAdelaide}
\affiliation{\instBonn}
\affiliation{\instUBC}
\affiliation{\instCincinnati}
\affiliation{\instFlorida}
\affiliation{\instHawaii}
\affiliation{\instHeidelberg}
\affiliation{\instLjubljanaUniLJ}
\affiliation{\instLouisville}
\affiliation{\instMalaya}
\affiliation{\instLjubljanaUM}
\affiliation{\instMelbourne}
\affiliation{\instMississippi}
\affiliation{\instUOM}
\affiliation{\instPittsburgh}
\affiliation{\instUSTC}
\affiliation{\instSAlabama}
\affiliation{\instSCarolina}
\affiliation{\instSydney}
\affiliation{\instUTokyo}
\affiliation{\instEri}
\affiliation{\instIPMU}
\affiliation{\instVictoria}
\affiliation{\instVPI}
\affiliation{\instWayneState}
\affiliation{\instYamagata}
\affiliation{\instYonsei}
  \author{F.~Abudin{\'e}n}\affiliation{\instTriesteINFN} 
  \author{I.~Adachi}\affiliation{\instKEK}\affiliation{\instSOKENDAI} 
  \author{R.~Adak}\affiliation{\instFudan} 
  \author{K.~Adamczyk}\affiliation{\instKrakow} 
  \author{L.~Aggarwal}\affiliation{\instPanjab} 
  \author{P.~Ahlburg}\affiliation{\instBonn} 
  \author{H.~Ahmed}\affiliation{\instXavier} 
  \author{J.~K.~Ahn}\affiliation{\instKoreaUnivKU} 
  \author{H.~Aihara}\affiliation{\instUTokyo} 
  \author{N.~Akopov}\affiliation{\instYerevan} 
% ...
\collaboration{Belle II Collaboration}

%% file: acknowledgements.tex
We thank the SuperKEKB group for the excellent operation of the
accelerator; the KEK cryogenics group for the efficient
operation of the solenoid; and the KEK computer group for
on-site computing support.
This work was supported by the following funding sources:
Science Committee of the Republic of Armenia Grant No. 18T-1C180;
Australian Research Council and research grant Nos.
DP180102629, 
DP170102389, 
DP170102204, 
DP150103061, 
FT130100303, 
and
FT130100018; 
Austrian Federal Ministry of Education, Science and Research, and
Austrian Science Fund No. P 31361-N36; 
Natural Sciences and Engineering Research Council of Canada, Compute Canada and CANARIE;
Chinese Academy of Sciences and research grant No. QYZDJ-SSW-SLH011,
National Natural Science Foundation of China and research grant Nos.
11521505,
11575017,
11675166,
11761141009,
11705209,
and
11975076,
LiaoNing Revitalization Talents Program under contract No. XLYC1807135,
Shanghai Municipal Science and Technology Committee under contract No. 19ZR1403000,
Shanghai Pujiang Program under Grant No. 18PJ1401000,
and the CAS Center for Excellence in Particle Physics (CCEPP);
the Ministry of Education, Youth and Sports of the Czech Republic under Contract No.~LTT17020 and 
Charles University grants SVV 260448 and GAUK 404316;
European Research Council, 7th Framework PIEF-GA-2013-622527, 
Horizon 2020 Marie Sklodowska-Curie grant agreement No. 700525 `NIOBE,' 
and
Horizon 2020 Marie Sklodowska-Curie RISE project JENNIFER2 grant agreement No. 822070 (European grants);
L'Institut National de Physique Nucl\'{e}aire et de Physique des Particules (IN2P3) du CNRS (France);
BMBF, DFG, HGF, MPG, AvH Foundation, and Deutsche Forschungsgemeinschaft (DFG) under Germany's Excellence Strategy -- EXC2121 ``Quantum Universe''' -- 390833306 (Germany);
Department of Atomic Energy and Department of Science and Technology (India);
Israel Science Foundation grant No. 2476/17
and
United States-Israel Binational Science Foundation grant No. 2016113;
Istituto Nazionale di Fisica Nucleare and the research grants BELLE2;
Japan Society for the Promotion of Science,  Grant-in-Aid for Scientific Research grant Nos.
16H03968, 
16H03993, 
16H06492,
16K05323, 
17H01133, 
17H05405, 
18K03621, 
18H03710, 
18H05226,
19H00682, 
26220706,
and
26400255,
the National Institute of Informatics, and Science Information NETwork 5 (SINET5), 
and
the Ministry of Education, Culture, Sports, Science, and Technology (MEXT) of Japan;  
National Research Foundation (NRF) of Korea Grant Nos.
2016R1\-D1A1B\-01010135,
2016R1\-D1A1B\-02012900,
2018R1\-A2B\-3003643,
2018R1\-A6A1A\-06024970,
2018R1\-D1A1B\-07047294,
2019K1\-A3A7A\-09033840,
and
2019R1\-I1A3A\-01058933,
Radiation Science Research Institute,
Foreign Large-size Research Facility Application Supporting project,
the Global Science Experimental Data Hub Center of the Korea Institute of Science and Technology Information
and
KREONET/GLORIAD;
Universiti Malaya RU grant, Akademi Sains Malaysia and Ministry of Education Malaysia;
Frontiers of Science Program contracts
FOINS-296,
CB-221329,
CB-236394,
CB-254409,
and
CB-180023, and SEP-CINVESTAV research grant 237 (Mexico);
the Polish Ministry of Science and Higher Education and the National Science Center;
the Ministry of Science and Higher Education of the Russian Federation,
Agreement 14.W03.31.0026;
University of Tabuk research grants
S-1440-0321, S-0256-1438, and S-0280-1439 (Saudi Arabia);
Slovenian Research Agency and research grant Nos.
J1-9124
and
P1-0135; 
Agencia Estatal de Investigacion, Spain grant Nos.
FPA2014-55613-P
and
FPA2017-84445-P,
and
CIDEGENT/2018/020 of Generalitat Valenciana;
Ministry of Science and Technology and research grant Nos.
MOST106-2112-M-002-005-MY3
and
MOST107-2119-M-002-035-MY3, 
and the Ministry of Education (Taiwan);
Thailand Center of Excellence in Physics;
TUBITAK ULAKBIM (Turkey);
Ministry of Education and Science of Ukraine;
the US National Science Foundation and research grant Nos.
PHY-1807007 
and
PHY-1913789, 
and the US Department of Energy and research grant Nos.
DE-AC06-76RLO1830, 
DE-SC0007983, 
DE-SC0009824, 
DE-SC0009973, 
DE-SC0010073, 
DE-SC0010118, 
DE-SC0010504, 
DE-SC0011784, 
DE-SC0012704; 
and
the National Foundation for Science and Technology Development (NAFOSTED) 
of Vietnam under contract No 103.99-2018.45.